\documentclass{amsart}

\usepackage[utf8]{inputenc}
\usepackage[T2A]{fontenc}
\usepackage{amsthm}
\usepackage{graphicx,textcomp}
\usepackage{amsmath,bm,amsfonts,mathrsfs,amssymb,bbm}
\usepackage{hyperref}
\newtheorem{theorem}{Theorem}[section]
\newtheorem{lemma}[theorem]{Lemma}

\newtheorem{rem}[theorem]{Remark}

\usepackage{appendix}
\usepackage[normalem]{ulem}
\usepackage{graphicx,color}
\usepackage{xcolor}

\begin{document}
\title[Dirichlet-to-Neumann invariant]{On the Dirichlet-to-Neumann conformal invariant of bounded planar domains}

\author{Alexey Kokotov}
	\address{Department of Mathematics \& Statistics, Concordia University, 1455 De Maisonneuve Blvd. W. Montreal, QC  H3G 1M8, Canada.  \url{https://orcid.org/0000-0003-1940-0306}}
	\email{alexey.kokotov@concordia.ca}

\author{Dmitrii Korikov}
	\address{St. Petersburg Department of Steklov Mathematical Institute
of Russian Academy of Sciences, 27 Fontanka, St. Petersburg, Russia. \url{https://orcid.org/0000-0002-3212-5874}}
	\email{dmitrii.v.korikov@gmail.com}
	
	\author{Marina Nenasheva}
	\address{Leonhard Euler International Mathematical Institute in Saint Petersburg, Pesochnaya nab. 10, St. Petersburg, 197022, Russia. \url{https://orcid.org/0000-0002-0508-8713}}
	\email{marina.nenasheva@skoltech.ru}

\begin{abstract}
Using  an analogue of the Mandelstam-Giddings-Wolpert diagrams, we introduce a new canonical representative of the conformal class of a bounded domain of arbitrary connectivity in $\mathbb{C}$ as a flat conical surface with geodesic boundary (a "truncated light-cone diagram",  simply LC-diagram in the sequel). The space of diagrams carries natural coordinates. We derive variational formulas for the determinant of the  Dirichlet Laplacian on a LC-diagram with respect to these coordinates. Then, passing to the Schottky double of the LC-diagram, making use of the Burghelea-Friedlander-Kappeler formula and the known variational formulas for determinants of Laplacians on the moduli space of holomorphic differentials, we compute the Dirichlet-to-Neumann (DN) conformal invariant $\frac{{\rm det}'\Lambda}{|\Gamma|}$ (here $\Lambda$ is the DN operator on the boundary, $\Gamma$, of a multiply connected domain and $|\Gamma|$ is the length of the boundary). The resulting formula (which uses the periods of the Schottky double of the domain only) provides an elementary counterpart to the formulas of Guillarmou and Guillop\'e who had expressed the DN invariant through the Ruelle and Selberg zeta-functions. Our formula agrees with the recent result of Wentworth on the asymptotics of the DN invariant as all but one  boundary components shrink; we have used this result to fix the undetermined constant of integration in our formula. 
As a corollary, we derive an explicit formula for the determinant of the Dirichlet Laplacian in a multiply connected domain.  
      
\end{abstract}

	\maketitle

\section{Introduction}

The study of the moduli space of planar multiply connected domains is a classical subject going back at least to Hilbert \cite{Hilbert}, who proved that any such domain is conformally equivalent to 
the complex plane with a finite number of straight slits parallel to a given direction. Later, the list of possible canonical representatives of the conformal class of a multiply connected domain was significantly extended by many authors (see, e.g., the list containing six canonical forms on pp. 89-90 in \cite{Bergman}). 

Relatively recently, motivated by the study of spectral invariants (the $\zeta$-regularized determinants of the Dirichlet Laplacian and the Dirichlet-to-Neumann operator),  Osgood, Phillips and Sarnak \cite{OPS} and (respectively)  Guillarmou and Guillop\'e \cite{Guillarmou} (the latter authors considered Riemann surfaces of arbitrary genus with boundary) used canonical models having the form of a flat domain with boundary of constant geodesic curvature and (respectively) a complete surface of infinite volume with metric of constant negative curvature (or a hyperbolic surface with geodesic boundary). 

In particular, in \cite{Guillarmou} it was shown that the conformal invariant  $\frac{{\rm det}'\Lambda}{|\Gamma|}$ (here $\Lambda$ is the DN operator on the boundary, $\Gamma$, of a compact Riemann surface with boundary and $|\Gamma|$ is the length of the boundary; ${\rm det}'$ denotes the zeta-regularized determinant with zero modes excluded) can be expressed through the length spectrum invariants (Ruelle and Selberg $\zeta$-functions) of the canonical models with hyperbolic metrics. On the other hand, in \cite{OPS} model domains with a flat metric and  boundaries of constant geodesic curvature were used to prove that the height function (minus the logarithm of the determinant of the  Dirichlet Laplacian) tends to infinity as the multiply  connected plane domain degenerates. 

In the present paper we introduce a new canonical model of a planar multiply connected domain: a surface obtained via gluing a finite number of cylinders (in the spirit of the well-known Mandelstam diagrams from the string theory). This is a flat surface with (smooth) geodesic boundary and a finite number of conical points. It turns out that the set of these canonical models can be nicely coordinatized and, as a certain addendum to the results of \cite{OPS}, we derive the variational formulas of the height function of the canonical model of a multiply connected domain with respect to the coordinates.

The main result of the paper (and the main application of the new canonical model) concerns the DN invariant, $\frac{{\rm det}'\Lambda}{|\Gamma|}$, of a multiply connected domain $\mathcal{D}\subset {\mathbb C}$. Quite unexpectedly (at least for the authors), it turns out that it admits a strikingly simple explicit expression through the matrix of the $b$-periods, ${\mathbb B}$,  of the Schottky double of the multiply connected domain (with respect to a special canonical basis of cycles on the double).
Namely, we prove that
  \begin{equation}\label{MAIN}\frac{{\rm det}'\Lambda}{|\Gamma|}=2^n{\rm det}\Im {\mathbb B}\,,\end{equation}
where $n+1\geq 2$ is the number of the boundary components. 
{
\begin{rem}\footnote{We thank D. Korotkin for suggesting that we include this remark.}
The above special canonical basis of cycles is obtained as follows: { the cycle $b_j$ ($j=1,\dots,n$) is the $j$-th connected component, $\Gamma_j$, of the ``interior'' part of the boundary $\partial\mathcal{D}$ while $a_j=l_j-\tau(\l_j)$, where $\tau$ is the anti-holomorphic involution and the chain $l_j$ connects the ``exterior'' part $\Gamma_0$ of $\partial\mathcal{D}$ with $\Gamma_j$ in such a way that  $l_j\cap l_i=\varnothing$ for $i\ne j$. This basis depends only on the ordering of the connected components of $\partial\mathcal{D}$.} 


 A reordering of the $a$-cycles (together with reordering of the corresponding $b$-cycles) does not change the determinant of the imaginary part of the matrix of $b$-periods, due to the formula
\begin{equation}\label{smena}{\rm det} \Im \tilde {\mathbb B}=|{\rm det}(C{\mathbb B}+D)|^{-2}{\rm det} \Im {\mathbb B}\end{equation}
(see, e.g., \cite{Siegel}, \$4.5),
where 
$$(\tilde a_1, \dots, \tilde a_{n}, \tilde b_1, \dots, \tilde b_{n})^t=
\left(\begin{matrix}D\ \  C\\ B\ \  A \end{matrix}\right)(a_1, \dots,  a_{n},  b_1, \dots,  b_{n})^t,$$
with $$\left(\begin{matrix}A\ \  B\\ C\ \  D \end{matrix}\right)\in {\rm Sp}(2n,{\mathbb Z})\,.$$
(For this reordering one has $C=0$ and $|{\rm det\,}D|=1$.)

Moreover, an elementary combinatorial topology reasoning (together with (\ref{smena}) and simple linear algebra) shows that the right-hand side of
 (\ref{MAIN}) does not change if  (passing from ${\mathbb C}$ to the Riemann sphere) one makes use as the "exterior" part of the boundary, 
 the boundary of one of the holes (say, $\Gamma_1$)
 and then repeats the above construction of a canonical basis on the double. { In this case, we have $C=0$ and
\begin{align*}
{\tiny 
 D=\left(\begin{array}{ccccc}
-1 & 0 & 0 & \dots & 0\\
-1 & 1 & 0 & \dots & 0\\
-1 \ & \ & \ & \dots & \ \\
-1 & 0 & 0 & \dots & 1
\end{array}\right).}
\end{align*}}
\end{rem}}

{
\begin{rem}
The matrix $2\Im\mathbb{B}$ coincides with capacitance matrix $\mathscr{C}$ with entries $\mathscr{C}_{ij}=\int_{\Gamma_i}\partial_\nu u_j dl$, $i, j=1, \dots, n$,
where $u_j$ is the harmonic function in $\mathcal{D}$ obeying $u_j=1$ on $\Gamma_j$ and $u_j=0$ on $\partial\mathcal{D}\backslash\Gamma_j$, and $\nu$ is the exterior normal vector. Thus, formula (\ref{MAIN}) can be rewritten as ${\rm det}'\Lambda/|\Gamma|={\rm det}\mathscr{C}$; in particular, this means that the DN invariant can be computed via effective numerical methods.
\end{rem}

}
 In particular, comparison of this expression with results of \cite{Guillarmou} leads to a relation between length spectrum invariants of a hyperbolic surface with boundary and the $b$-periods of its double. 
 
Moreover, as an immediate corollary of our result (together with a conical version of the Alvarez formula, the BFK formula and the results of \cite{KoKorot}), we derive an explicit formula for the determinant of the Dirichlet Laplacian in an arbitrary multiply connected domain in ${\mathbb C}$.  
 
It should be mentioned that our result allows two cross-checks. One can easily compare it with the case of surfaces of Euler characteristic zero from \cite{Guillarmou} (here a direct simple computation is possible, see Appendix to \cite{Guillarmou}) and with Wentworth's highly non-trivial computation (see \cite{Wentworth}, Theorem 3.1)  of the asymptotics of the DN invariant as all the components of the boundary except one shrink.

We note that our result can be extended to the case of compact Riemann surfaces of arbitrary genus with boundary having $n+1$ connected components (with $n\geq 1$!). To this end one should make use of truncated Mandelstam diagrams of higher genus; this will be done elsewhere.  We discuss the planar domains (and genus zero ("tree") diagrams) for the sake of brevity, this significantly reduces the amount of the needed pure technical work.  

We also note that the methods of the present paper {\it do not work} for Riemann surfaces with boundary having only one connected component (there are no Mandelstam diagrams with one pole, i.e., with one semi-infinite cylinder; see \cite{GiWolp}). Fortunately, in this case a different approach is possible (see our recent preprint \cite{KK}), which leads to a more complicated but essentially similar result.

Finally, we state that we are greatly indebted to \cite{Guillarmou}. Paradoxically, our main debt is not the statement of the problem we are solving here or important technical tricks we borrowed
from \cite{Guillarmou} (e. g., exploiting the conformal invariance of  $\frac{{\rm det}'\Lambda}{|\Gamma|}$, considering the Schottky double,  making use of the BFK formula). It is a minor remark of the authors of \cite{Guillarmou} made only in passing, where they stated that making use of flat metrics "does not seem to be apparent way to express the determinant of DN operator in terms of geometric invariants". This remark has provided us with a crucial idea and the main motivation.

\section{Moduli space of domains in $\mathbb{C}$ with multicomponent boundaries} 

Let $\mathcal{D}$ be a domain in $\mathbb{C}$ bounded by smooth simple closed curves $\Gamma_0,\Gamma_1,\dots,\Gamma_n$ ($n\ge 1$). We assume that $\Gamma_0$ coincides with the external part $\partial\mathcal{D}_{ext}:=\partial(\mathbb{C}\backslash\mathcal{D})$ of the boundary of $\mathcal{D}$. Let us write $\mathcal{D}\sim\mathcal{D'}$ if there is a biholomorphism $\beta:\,\mathcal{D}\to\mathcal{D}'$ such that $\beta$ and $\beta^{-1}$ are smooth up to the boundaries and $\beta(\partial\mathcal{D}_{ext})=\partial\mathcal{D}'_{ext}$.

In this paragraph, we introduce coordinates on the space $\mathfrak{D}_n$ of classes $[\mathcal{D}]$ under the above conformal equivalence. To this end, we construct the canonical light-cone model of $\mathcal{D}$. 

\subsection{Light-cone metrics.} Recall (see \cite{GiWolp}) that a {\it light-cone diagram} $\mathsf{M}_\infty$ is a non-compact surface obtained by gluing finite and semi-infinite cylinders in such a way that $\mathsf{M}_\infty$ is locally isometric to $\mathbb{C}$ or a cone of angle $2\pi k$ ($k=2,3,\dots$) and the number of conical points is finite. Each such $\mathsf{M}_\infty$ admits a time coordinate $t$ which is smooth outside the conical points and coincides, up to a constant, with a longitudinal coordinate on each cylinder.
\begin{lemma}
\label{domais vs reduced LC}
For each $\mathcal{D}$ there is a LC-diagram $\mathsf{M}_\infty$ and the smooth {\rm(}up to the boundaries{\rm)} biholomorphism from $\mathcal{D}$ onto the domain $t^{-1}((0,1))\subset\mathsf{M}_\infty$. Under the condition that $\mathsf{M}_\infty$ contains no conical points outside $t^{-1}((0,1))$, the LC-diagram $\mathsf{M}_\infty$ depends only on $[\mathcal{D}]$ and the maps $[\mathcal{D}]\mapsto\mathsf{M}_\infty$ and $\mathfrak{A}:\,[\mathcal{D}]\mapsto\mathsf{M}:=\{x\in\mathsf{M}_\infty \ | \ t(x)\in[0,1] \}$ are injective.
\end{lemma}
\begin{proof}
Let $u$ be a harmonic function on $\mathcal{D}$ equal to 0 on $\partial\mathcal{D}_{ext}$ and to 1 on $\partial\mathcal{D}\backslash\partial\mathcal{D}_{ext}:=\partial\mathcal{D}_{int}$. Introduce the holomorphic differential 
\begin{equation}
\label{abel diff}
\omega:=du+i\star du=2\partial u
\end{equation}
(where $\star$ is the standard Hodge star operator on $\mathbb{C}$) and endow $\mathcal{D}$ with the flat conical conformal metric 
$$m:=|\omega|^2.$$ 
Then the (multivalued) function 
\begin{equation}
\label{flat coordinate}
x\mapsto z(x):=\int_\cdot^x\omega
\end{equation}
provides a smooth isometric embedding of each simply connected domain in $\mathcal{D}$ containing no zeroes of $m$ into $\mathbb{C}$; if $x_0\in \mathcal{D}$ is a zero of $du$ of order $k$, then (\ref{flat coordinate}) provides an isometric embedding of a small neighborhood $x_0$ into the Riemann surface of $z\mapsto\sqrt[k+1]{z}$. 

The level and gradient lines of $u$ are geodesics in the metric $m$ (outside the zeroes of $du$). Thus, if $c$ is not a critical value of $u$, then there are the geodesic parallel coordinates near $u^{-1}(\{c\})$ which can be extended to any domain $u^{-1}((a,b))$, where $(a,b)\ni c$ contains no critical values of $u$ (indeed, each geodesic orthogonal to $u^{-1}(\{c\})$ is a gradient line of $u$; gradient lines can intersect only at the zeroes of $du$). This means that the surface $(u^{-1}((a,b)),m)$ is isometric to the interior of $[a,b]\times(u^{-1}(\{c\}),dl_m)$, where $dl_m$ is the length element induced by $m$ on $u^{-1}(\{c\})$. In particular, a connected component $V$ of $(u^{-1}((a,b)),m)$ is isometric to the interior of the cylinder $\Pi=[a,b]\times(\mathbb{R}/T(V)\mathbb{Z})$, where $T(V):=-i\int_{u^{-1}(\{c\})\cap V}\omega$; the isometry being provided by the map 
\begin{equation}
\label{cyliner isometry}
x\mapsto (\Re z(x),\Im z(x)/T(V)\mathbb{Z}),
\end{equation}
where $z$ is given by (\ref{flat coordinate}). Since map (\ref{flat coordinate}) is a local diffeomorphism outside the zeroes of $du$, the above isometries and their inverses are smooth up to the boundaries except the zeroes of $du$ and their images. 

Now the gluing lemma for holomorphic functions and the theorem on removable singularities yields that local maps (\ref{cyliner isometry}) can be glued together to provide an isometry $\beta$ from $(\mathcal{D},m)$ onto the interior of the surface $\mathsf{M}$ glued from the above cylinders $\Pi$ (where $a,b$ are critical values of $u$) which is smooth up to $\partial\mathcal{D}$. To prove that $\beta^{-1}$ is smooth up to $\partial\mathsf{M}$, it remains to show that $du$ has no zeroes on $\partial\mathcal{D}$; the latter follows from the normal derivative lemma for harmonic functions.

Finally, attaching the cylinders $(-\infty,0]\times\beta(\Gamma_0)$, $[1,+\infty)\times\beta(\Gamma_1),\dots,[1,+\infty)\times\beta(\Gamma_n)$ to $\beta(\Gamma_0)$, $\beta(\Gamma_1),\dots,\beta(\Gamma_n)$, respectively, one obtains the LC-diagram $\mathsf{M}_\infty$ while the function $u$ can be continued to a time coordinate $t$ on $\mathsf{M}_\infty$ obeying $t\circ\beta=u$. (Conversely, the reasoning before (\ref{cyliner isometry}) shows that if $\mathsf{M}_\infty$ contains no conical points outside $t^{-1}((0,1))$, then $\mathsf{M}_\infty\backslash t^{-1}((0,1))$ is isometric to $\partial\mathsf{M}\times[0,+\infty)$.) Since the conformal maps preserve the harmonicity, $\mathsf{M}_\infty$ depends only on $[\mathcal{D}]$. If the above construction provides the same LC-diagram for the surface $\mathcal{D}'$, then $(\mathcal{D}',m')$ is isometric (thus, biholomorphic) to $(\mathcal{D},m)$. The external parts $\Gamma_0=\partial\mathcal{D}_{ext}$, $\Gamma'_0=\partial\mathcal{D}'_{ext}$ have maximal lengths (in the metrics $m$, $m'$, resp.) among connected components of $\partial\mathcal{D}$, $\partial\mathcal{D}'$ due to the Green formula
$$|\Gamma_0|_m=\int_{\partial\mathcal{D}_{ext}}\partial_\nu udl=-\sum_{k=1}^n\int_{\Gamma_k}\partial_\nu udl=\sum_{k=1}^n|\Gamma_k|_m,$$
Thus, the above isometry maps $\partial\mathcal{D}'_{ext}$ onto $\partial\mathcal{D}_{ext}$ which implies $\mathcal{D}\sim\mathcal{D}'$.
\end{proof}
\begin{figure}[h!]
\center{\includegraphics[width=1\linewidth]{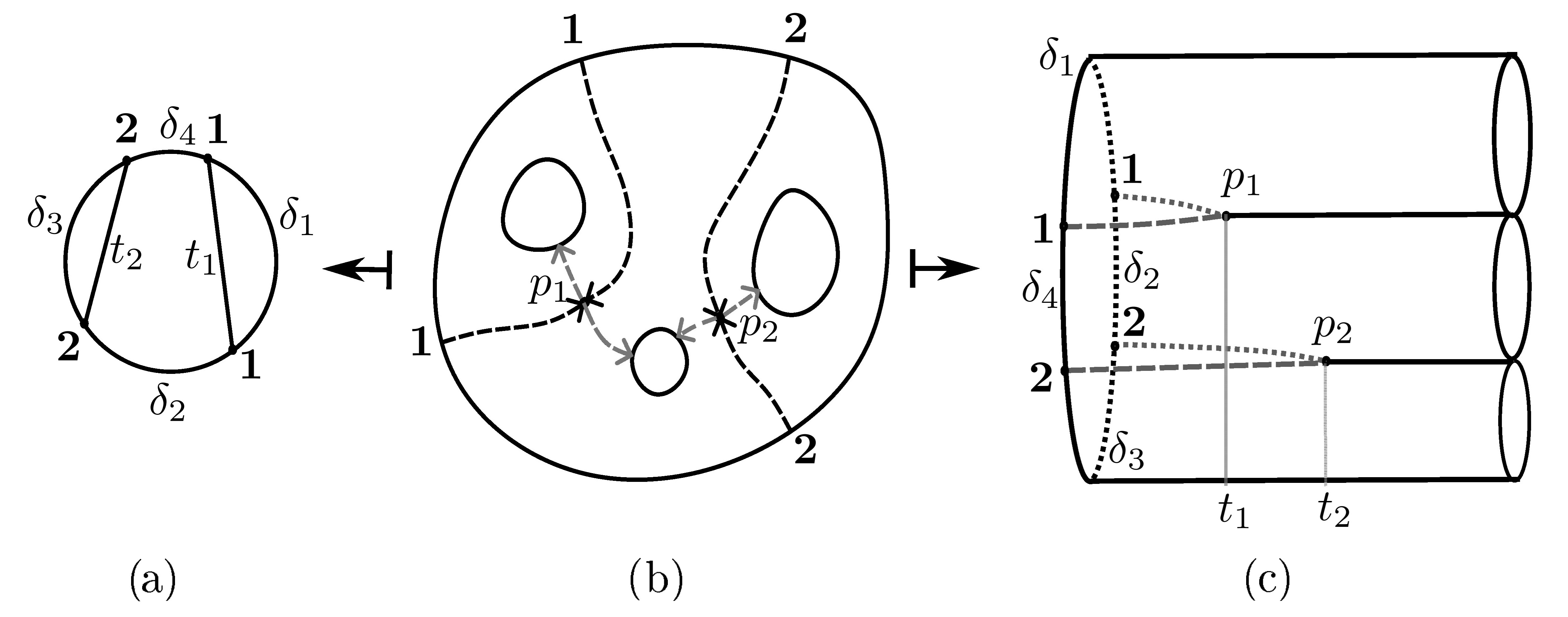}}
\caption{The domain $\mathcal{D}$ with the gradient lines meeting at critical points of $dt$ (b), its diagram $\mathsf{M}=\mathfrak{A}([\mathcal{D}])$ (c), and the associated NC partition (a).}
\label{NCLC}
\end{figure}
In what follows, the surfaces $\mathsf{M}$ constructed in Lemma \ref{domais vs reduced LC} are called diagrams. 

\subsection{Coordinates.} Now we give a coordinatization of top-dimensional cells in $\mathfrak{D}_n$. Let $n>1$. Suppose that all the conical points $p_k$ of $\mathsf{M}$ are of angle $4\pi$ (that is, all zeroes of $dt$ are simple); then their total number is $n-1$ due to the Gauss-Bonnet theorem and the fact that the boundary $\partial\mathsf{M}$ is geodesic. In addition, suppose that there are no gradient or level lines of $t=u$ joining two conical points. We associate with $\mathsf{M}=\mathfrak{A}([\mathcal{D}])$ the weighted non-crossing (NC) partition constructed as follows (see Fig. \ref{NCLC}). Let $\mathcal{D}_\circ$ be a domain bounded by $\partial\mathcal{D}_{ext}$ (via homeomorphism, one can identify $\mathcal{D}_\circ$ with the unit circle). We construct the circular graph on $\mathcal{D}_\circ$ whose internal edges are constituted by the gradient lines of $u=t$ incoming to the zeroes of $dt=du$, while the boundary vertices are the starting points of these lines. We mark the internal edge by the number ${\bf k}$ (and group together its ends) if it is formed by the gradient lines meeting at the same zero $p_k$ of $dt$ (since the gradient lines cannot intersect outside conical points, the obtained partition of $\partial\mathcal{D}_{ext}\equiv\mathbb{T}$ is NC by construction). To each internal edge ${\bf k}$, we assign the weight 
$$t_k:=t(p_k).$$ 
Each boundary arc $\mathscr{A}_s$ ($s\in\mathbb{Z}\big/2(n-1)\mathbb{Z}$) is clockwise-oriented, numbered in the clockwise order, and endowed with the weight 
$$\delta_s:=\int_{\mathscr{A}_s}\star dt.$$ 
Note that all the gradient lines of $t$ cut the diagram $\mathsf{M}$ into the rectangles $[0,1]\times[0,\delta_s]$, where each edge $[0,1]\times\{0\}$, $[0,1]\times\{\delta_s\}$ contains exactly one conical point. 

\begin{figure}
\center{\includegraphics[width=0.9\linewidth]{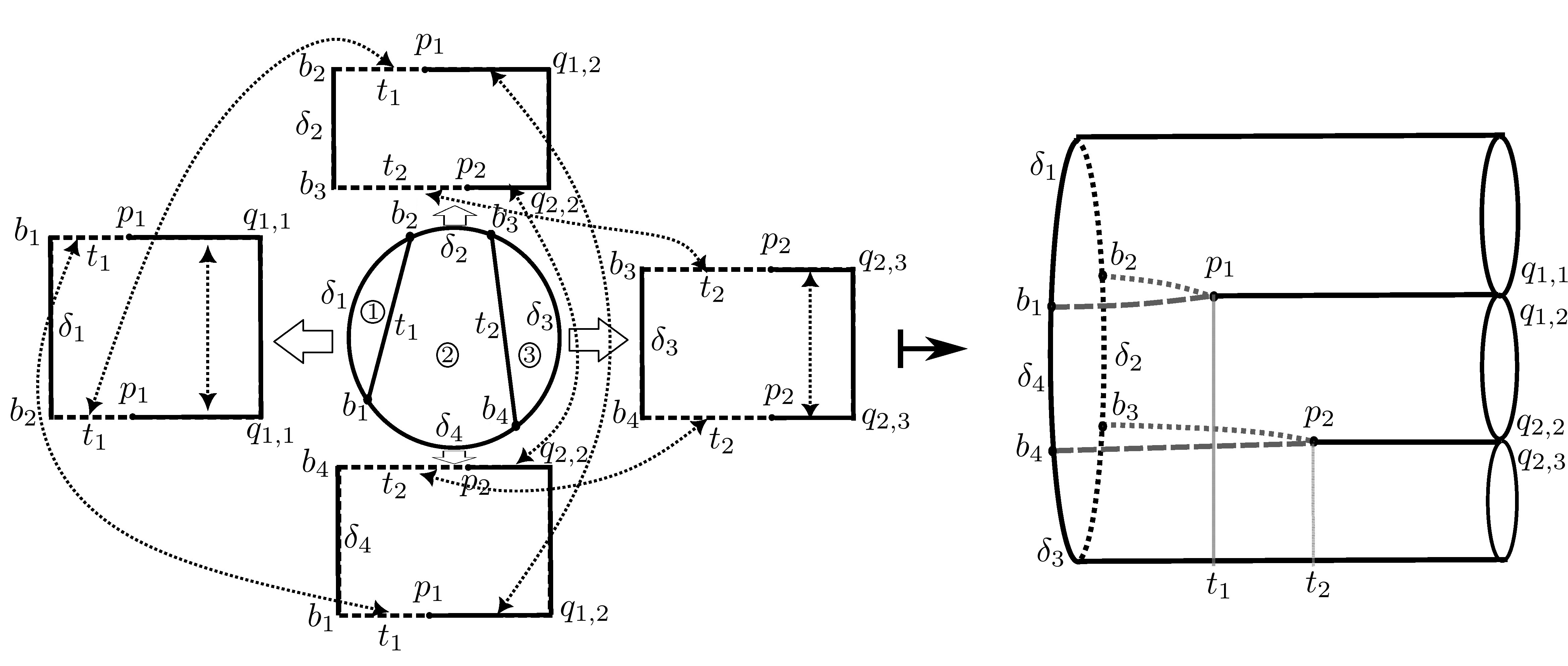}}
\caption{Assembling $\mathsf{M}$ from rectangles via its NC-partition.}
\label{Assembling}
\end{figure}

Thus, given the NC partition of $\mathsf{M}$, one can reconstruct the diagram $\mathsf{M}$ as follows. The edges of the NC partition divide the disk into connected components $\mathbb{D}_l$ ($l=1,\dots,n$) associated with the connected components $\Gamma_l$ of $\partial\mathcal{D}\backslash\partial\mathcal{D}_{ext}$. With the boundary arc $\mathscr{A}_s$ of weight $\delta_s$ and with the starting and ending points $b_s\in{\bf k}$ and $b_{s+1}\in{\bf j}$, respectively, we associate the rectangle $\Pi_s:=[0,1]\times[0,\delta_s]$ with six marked points
\begin{align*}
b_{s}\equiv\{0\}\times\{0\}, \quad b_{s+1}\equiv\{\delta_s\}\times\{0\},\\
p_k\equiv\{0\}\times\{t_k\}, \quad p_j\equiv\{\delta_s\}\times\{t_j\},\\
q_{k,l}\equiv\{0\}\times\{1\}, \quad q_{j,l}\equiv\{\delta_s\}\times\{1\}.
\end{align*}
where $\mathscr{A}_s\subset\partial\mathbb{D}_l$. Gluing the above rectangles along all their coinciding boundary segments $[b_{s},p_k]$, $[p_k,q_{k,l}]$ provides the required diagram $\mathsf{M}$ (see Fig. \ref{Assembling}).
So, the numbers 
\begin{equation}
\label{coordinates diagram}
t_1,\dots,t_{n-1}, \quad \delta_1,\dots,\delta_{2(n-1)}
\end{equation} 
provide the local coordinates of $[\mathcal{D}]$ on the top-dimensional cell $\mathscr{C}\equiv (0,1)^{n-1}\times(0,+\infty)^{2(n-1)}$ in the moduli space $\mathfrak{D}_n$ while the NC partition itself is a combinatorial parameter that determines $\mathscr{C}$ itself.

If $n=1$, then (\ref{flat coordinate}) maps $\mathcal{D}$ onto a cylinder $[0,1]\times(\mathbb{R}/\delta\mathbb{Z})$, in this case, the number $\delta>0$ is the unique coordinate of $[\mathcal{D}]$.

\subsection{Deformations of light-cone diagrams corresponding to small variations of coordinates.} 
\begin{figure}[h!]
\center{\includegraphics[width=0.8\linewidth]{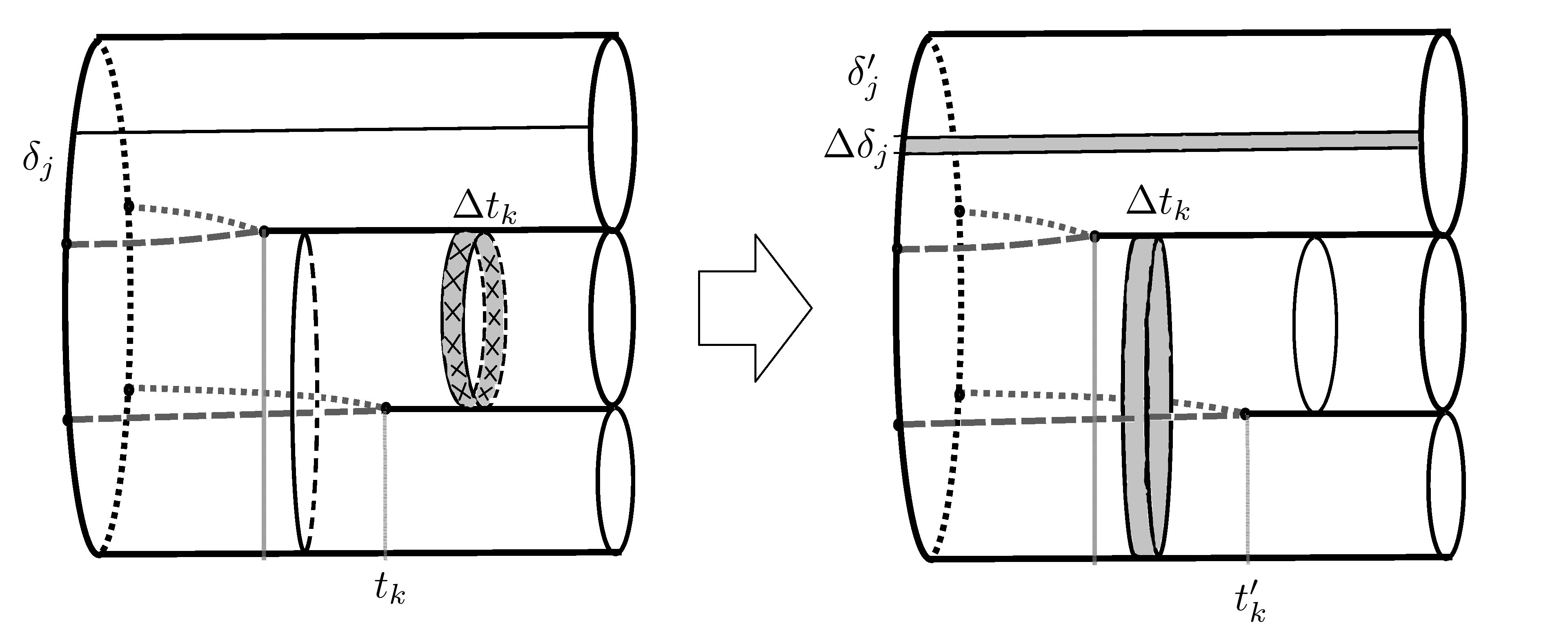}}
\caption{Gluing in/deleting the rectangles/rings corresponding to the variations $t_k\mapsto t'_k:=t_k+\Delta t_k$ and $\delta_s\mapsto\delta'_s:=\delta_s+\Delta\delta_s$ of the coordinates of $\mathsf{M}$.}
\label{variation}
\end{figure}

Small variations of the above coordinates can be realized by gluing in or deleting small horizontal strips or vertical rings, as shown in Fig. \ref{variation}. 

\begin{figure}[h!]
\center{\includegraphics[width=1\linewidth]{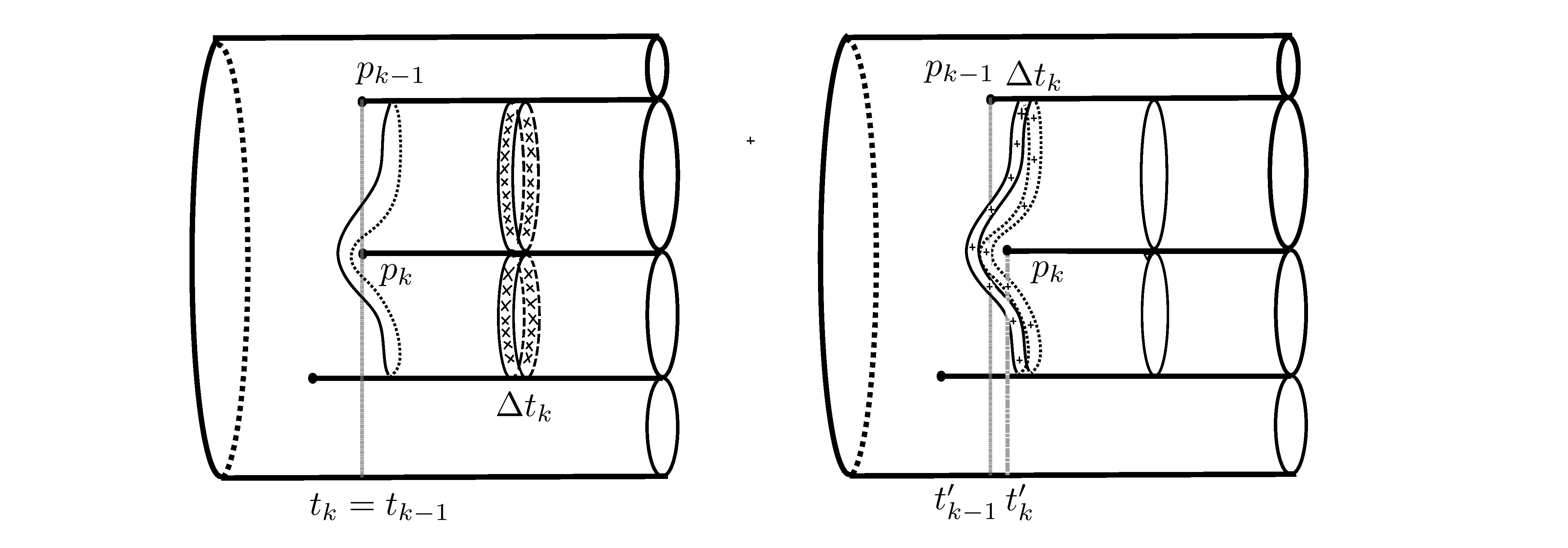}}
\caption{Deformation $t_k\mapsto t'_k:=t_k+\Delta t_k$ of the diagram obeying $t_{k}=t_{k-1}$.}
\label{treordering}
\end{figure}

\begin{figure}[h!]
\center{\includegraphics[width=1\linewidth]{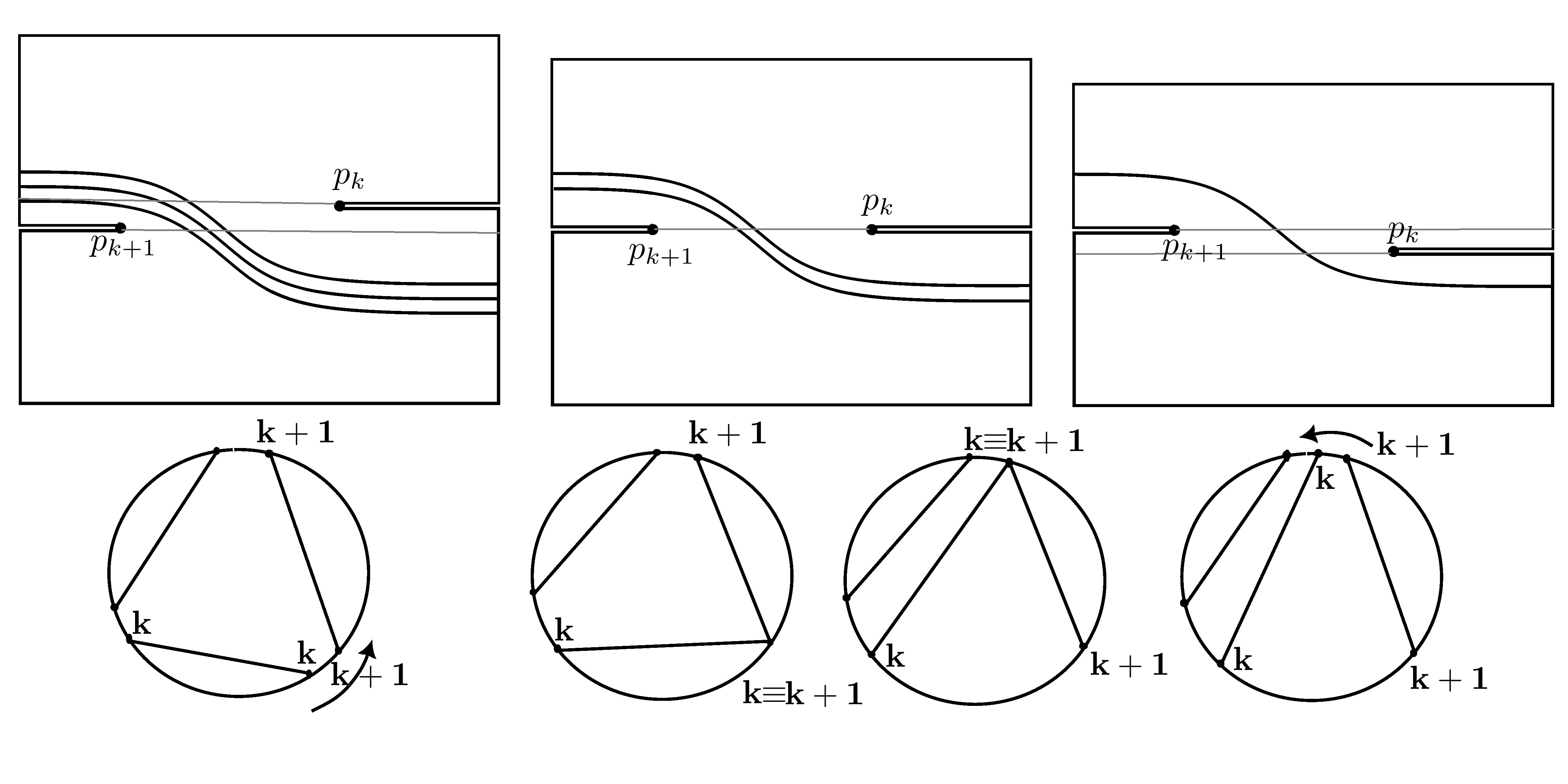}}
\caption{Deformations of $\mathsf{M}$ with the gradient line of $t$ joining two conical points $p_k$, $p_{k+1}$ shifting $[\mathcal{D}]$ to the interiors of the top-dimensional cells (top) corresponding to different NC partitions (bottom). Only strips containing $p_k$, $p_{k+1}$ are shown.}
\label{NCreordering}
\end{figure}

To identify the boundaries of the top-dimensional cells $\mathscr{C}$ of $\mathfrak{D}_n$ corresponding to different partitions $q$ of $\mathbb{T}$, one also needs to consider small deformations of the diagrams containing conical points connected by the level or gradient line of $t$. Note that, in this case, the gradient lines of $t$ outgoing from/ingoing to the conical points still cut the diagram $\mathsf{M}$ into several rectangles, but their horizontal sides may contain several conical points. If the conical points $p_k$, $p_{k-1}$ are connected by the level line of $t$, then the deformation $t_k\mapsto t'_k:=t_k+\Delta t_k$ is performed by gluing in/deleting thin distorted rings bounded by geodesically parallel curves (see Fig. \ref{treordering}) and shifts the class $[\mathcal{D}]$ to the interior of the top-dimensional cell. 
Similarly, the variation shown in Fig. \ref{NCreordering} and performed by gluing in/deleting a thin strip bounded by geodesically parallel curves leads to the reconfiguration of the NC partition. Note that applying  the above reconfigurations subsequently, one can transform any NC partition to the trivial one, which is the disk dissected by parallel edges. 

Therefore, the above variations allow one to glue all the above top-dimensional cells $\mathscr{C}\equiv (0,1)^{n-1}\times(0,+\infty)^{2n-2}$ in $\mathfrak{D}_n$ together to provide the (connected) stratum $\acute{\mathfrak{D}}_n\subset\mathfrak{D}$ of moduli of domains $\mathcal{D}$ whose differentials $dt$ have simple zeroes.

\subsection{Schottky embedding.} 
\begin{figure}[h!]
\center{\includegraphics[width=1\linewidth]{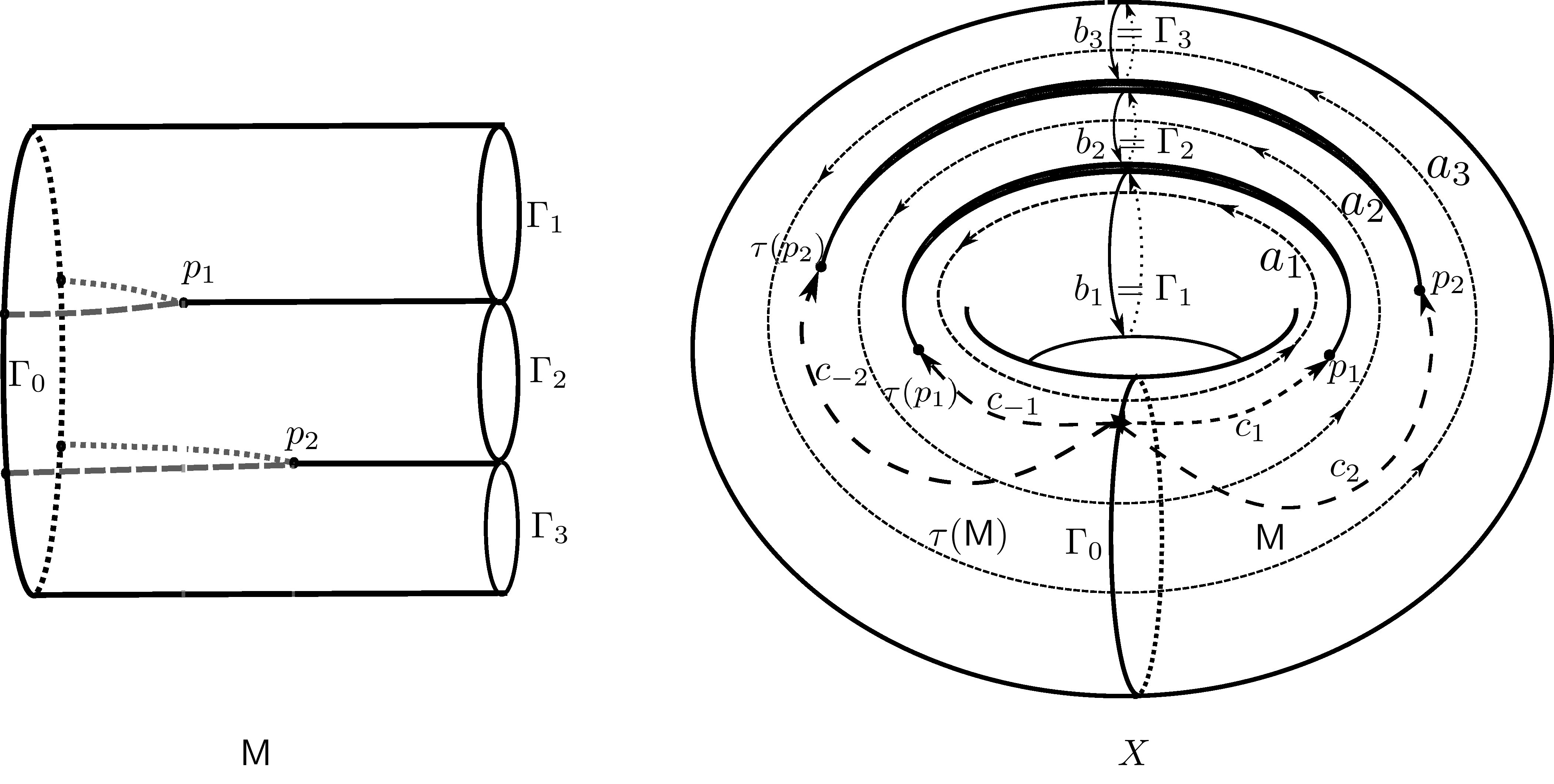}}
\caption{The diagram $\mathsf{M}$ and its double $X$. The contours $a_i$, $b_j=\Gamma_j$ and $c_{\pm k}$ are shown on $X$.}
\label{double}
\end{figure}

For the domain $\mathcal{D}$, introduce its Schottky double $(X,\omega,\tau)=:\mathfrak{S}([\mathcal{D}])$ which is the genus $n$ Riemann surface obtained by gluing $\mathsf{M}=\mathfrak{U}([\mathcal{D}])$ with its copy (endowed with opposite orientation) along the boundaries (see Fig. \ref{double}); here $\tau$ is the anti-holomorphic involution on $X$ interchanging the points of the $\mathsf{M}\subset X$ and its copy and the holomorphic differential $\omega$ is obtained by the extension of differential (\ref{abel diff}) from $\mathsf{M}$ onto $X$ by anti-symmetry, 
\begin{equation}
\label{anti-symmetry of differential}
\tau^*\omega=-\overline{\omega}.
\end{equation}
Thus, the map $\mathfrak{S}:\,[\mathcal{D}]\mapsto(X,\omega,\tau)$ provides a (ramified) cover of $\acute{D}_n$ over the real locus of the stratum $H_n(1,\dots,1)$ in the moduli space of Abelian differentials consisting of differentials with simple zeroes (note that $\mathfrak{S}$ is not injective due to the presence of higher symmetric doubles: for example, for $n=2$, each $X$ is hyperelliptic and admits two involutions the quotients by which are not, in general, conformally equivalent).

The canonical local coordinates on $H_n(1,\dots,1)$ are given by the periods of the Abelian differentials with respect to the relative homology basis specified on the base surface (see, e.g., \cite{KoKorot}). Namely, if $\omega$ is a holomorphic differential with simple zeroes on a Riemann surface $X$, then the coordinates of $(X,\omega)$ are given by
\begin{equation}
\label{moduli coordinates}
\begin{split}
A_i&=\int_{a_i}\omega, \quad B_i=\int_{b_i}\omega \quad (i=1,\dots,n),\\ \tilde{C}_k&=C_k-C_1 \quad \Big(\,i=2,\dots,2(n-1), \qquad C_k=\int_{c_k}\omega\,\Big)
\end{split}
\end{equation}
where $a_i,b_i$ constitute the canonical homology basis on $X$ and each contour $c_k$ connects some base point with $k$-th zero of $\omega$. For the doubles of $(X,\omega,\tau)=:\mathfrak{S}([\mathcal{D}])$, these coordinates take the form
\begin{align*}
A_i&=\int_{a_i}\omega=1, \qquad B_i=\int_{b_i}\omega=i\sum_s\delta_s;\\
C_{\pm k}&=-\overline{C_{\mp k}}=\int_{c_{\pm k}}\omega=\pm t_k+i\sum_s\delta_s,
\end{align*}
where the homology basis $\{a_i,b_i\}_{i=1}^n$ on $(X,\omega,\tau)$ is chosen in such a way that $a_i$ and $b_i$ are real and imaginary curves of the differential $\omega$, respectively, and each curve $c_{\pm k}=\tau\circ c_{\mp k}$ connect the conical point $p_{\pm k}$ with the corner $b_1$ of the rectangle $\Pi_1$ (note that $b_i=\Gamma_i$, see Fig. \ref{double}). The summation in the first formula is taken over all $s$ obeying $\Pi_s\cap\Gamma_i\ne\varnothing$, where $\Pi_s$ are rectangles constituting $\mathsf{M}$ (i.e., over all $\delta_s$ such that the corresponding arcs in the NC partition belong to the boundary of the same connected component $\mathbb{D}_{i}$ of the disk). Similarly, the summation in the second formula is taken over all $\delta_s$ such that $\Pi_s$ intersect a patch $\mathfrak{P}_k\subset\Gamma_0$ connecting $b_1$ with the beginning of the gradient line of $t$ with end at $p_k$ (i.e., over such $\delta_s$ that the union of the corresponding boundary arcs in the NC partition connect a vertex of edge ${\bf 1}$ with a vertex of edge ${\bf k}$). As a corollary, we have
\begin{align}
\label{derivatives}
\begin{split}
\frac{\partial A_i}{\partial t_k}&=\frac{\partial A_i}{\partial \delta_j}=\frac{\partial B_i}{\partial t_k}=\frac{C_{\pm 1}}{\partial\delta_s}=0, \qquad \frac{\partial C_{\pm k}}{\partial t_{l}}=\pm\delta_{k,l},\\
\frac{\partial B_i}{\partial \delta_s}&=i\mathbbm{1}_\cap(\Pi_s,\Gamma_i), \qquad \frac{\partial C_{\pm k}}{\partial \delta_s}=i\mathbbm{1}_\cap(\Pi_s,\mathfrak{P}_k),
\end{split}
\end{align}
where the intersection indicator $\mathbbm{1}_\cap$ is given by
$$\mathbbm{1}_\cap(Q_1,Q_2)=\left\{\begin{array}{lr}
0 & Q_1\cap Q_2=\varnothing\\
\pm 1 & Q_1\cap Q_2\ne\varnothing
\end{array}\right. .$$

The basis in $H^0(X,K)$ dual to the homology basis $\{a_i,b_i\}_{i=1}^n$ is constructed as follows. Let $u_k$ ($k=0,\dots,n$) be a harmonic function $\mathcal{D}\equiv\mathsf{M}$ obeying $u_k=1$ on $\Gamma_k$ and $u_k=0$ on $\partial\mathcal{D}\backslash\Gamma_k$ (here the domain $\mathcal{D}$ and the diagram $\mathsf{M}$ are identified via the conformal equivalence provided by Lemma \ref{domais vs reduced LC}). Then the forms $\partial u_k=(du_k+i\star du_k)/2$ admit analytic continuation to the holomorphic differentials $\omega_k$ on the double $X\supset\mathsf{M}\equiv\mathcal{D}$ by anti-symmetry (\ref{anti-symmetry of differential}). By construction, these differentials obey $\int_{a_j}\omega_k=\delta_{jk}$ for $j,k>0$. Hence, the $b$-period matrix of $X$ is given by 
\begin{equation}
\label{capacitances}
2i\,\mathbb{B}_{ij}=2i\int_{b_i}\omega_j=-\int_{\Gamma_i}*du_j=-\int_{\Gamma_i}\partial_\nu u_j dl=:-\mathscr{C}_{ij}
\end{equation}
(here the right-hand side is independent of the choice of the conformal metric on $\mathsf{M}\equiv\mathcal{D}$). The matrix $\mathscr{C}=\{\mathscr{C}_{ij}\}_{i,j=1}^n$ is called the {\it capacitance matrix} of $\mathcal{D}$ (relative to the grounded exterior part $\Gamma_0$).
As a corollary, we have
\begin{equation}
\label{periods via capacities}
{\rm det}(\Im\mathbb{B})=2^{-n}{\rm det}(\mathscr{C}).
\end{equation}

\section{Variations of resolvent kernels, eigenvalues and zeta functions}
\subsection{Variations of resolvent kernels.} Let $(X,\omega)$ be a genus $n$ Riemann surface endowed with the holomorphic differential $\omega$ with simple zeroes and the corresponding flat conical metric $|\omega|^2$. We allow $(X,\omega)$ to depend on the parameter $\sigma$; for $\sigma$ one may choose $A_i$,$B_j$, of $\tilde{C}_k$ keeping constant all other coordinates (\ref{moduli coordinates}). The deformation of $(X,\omega)$ corresponding to the small variation of $\sigma$ can be performed via geometric surgery near the contours $A_i^\ddag=b_i$, $B_i^\ddag=-a_i$ and the small clockwise circle $C_k^\ddag=\tilde{C}_k^\ddag$ enclosing the $k$-th zero of $\omega$ (see \S 6.1, \cite{KoKori}). For example, if $\sigma=A_i,B_i$, then one removes from $(X,\omega)$ a thin neighborhood swept out under the parallel translation $\sigma^\ddag$ and then glues the boundaries or, conversely, cuts $(X,\omega)$ along $\sigma^\ddag$ and then glues in a thin closed strip. So, if the domain $\tilde{X}_\sigma$ does not intersect small neighborhoods of the above contours used in the surgery, then it can be considered as a common domain of all $(X,\omega)$ corresponding to  small variations of $\sigma$ and, for any function $(x,\sigma)\mapsto f(x,\sigma)$ on $X$, the derivative $\dot{f}(x,\sigma)=\frac{\partial f(x,\sigma)}{\partial\sigma}$ is well-defined on $\tilde{X}_\sigma$ but it depends on the choice of $\tilde{X}_\sigma$ (by construction, $\dot{\omega}=0$ on $\tilde{X}_\sigma$).

Let $R_\lambda$ be the resolvent kernel of the (non-negative Friedrichs) Laplacian $\Delta_{X}$ on $(X,|\omega|^2)$. According to Proposition 2, \cite{KoKorot}, the formula
\begin{equation}
\label{variation of resolvent kernel - formula}
\dot{R}_\lambda(x,y)=-2i\oint\limits_{\sigma^\ddag}\mathscr{W}_\lambda(x|\cdot|y) \qquad (\sigma=A_i,B_i,\tilde{C}_k, \quad x,y\in\tilde{X}_\sigma)
\end{equation}
is valid, where 
\begin{equation}
\label{variation of resolvent kernel - difform}
\mathscr{W}_\lambda(x|z|y)=\frac{\partial_z R_\lambda(x,z)\partial_z R_{\lambda}(z,y)-(\lambda/4)R_\lambda(x,z)R_\lambda(z,y)
|\omega(z)|^2}{\omega(z)}.
\end{equation}
Since $(\Delta_{X}-\lambda)R_\lambda(\cdot,z)=0$ outside $z$, form (\ref{variation of resolvent kernel - difform}) is closed. Thus, derivative (\ref{variation of resolvent kernel - formula}) is independent of the choice of the contour $\sigma^\ddag$ near which the geometric surgery is performed as long as it lies in the complement $X\backslash\tilde{X}_\sigma$ and does not change its homotopy class there (note that $\dot{R}_\lambda(x,y)$ has a jump along $\sigma^\ddag$ if $\tilde{X}_\sigma=X\backslash\sigma^\ddag$).

Now, let $\mathsf{M}=\mathfrak{A}([\mathbb{D}])$ be a diagram depending on the parameter $\sigma=t_k,\delta_j$ in such a way that all other coordinates (\ref{coordinates diagram}) are kept constant. Let $(X,\omega,\tau)=\mathfrak{S}([\mathcal{D}])$ be the Schottky double of $\mathsf{M}$. In this case, the deformation of $X$ corresponding to the small variation of $\sigma$ is performed by gluing in/deleting thin rings obtained by doubling thin rings/stripes used for deformation of the diagram $\mathsf{M}$ (see Fig. \ref{variation}). 

In view of (\ref{anti-symmetry of differential}), the involution $\tau$ is isometric on $(X,|\omega|^2)$ whence $\Delta_{X}(f\circ\tau)=(\Delta_{X}f)\circ\tau$. In particular, the resolvent kernel $R_\lambda$ is symmetric
\begin{equation}
\label{reslovent kernel symmetry}
R_\lambda(\tau(x),\tau(y))=R_\lambda(x,y).
\end{equation}
Since the boundary $\partial\mathsf{M}$ of $\mathsf{M}$ (embedded in $X$) coincides with the set of the fixed points of $\tau$, the function
\begin{equation}
\label{Dirichlet resolvet kernel}
\begin{split}
R^D_\lambda(x,y):=\frac{1}{2}\Big(R_\lambda(x,y)+R_\lambda(\tau(x),\tau(y))-R_\lambda(\tau(x),y)-R_\lambda(x,\tau(y))\Big)=\\
=R_\lambda(x,y)-R_\lambda(x,\tau(y))=-R_\lambda^D(x,\tau(y))
\end{split}
\end{equation}
obeys the Dirichlet conditions on $\partial\mathsf{M}$ and therefore its restriction on $\mathsf{M}$ coincides with the resolvent kernel of the Dirichlet Laplacian $\Delta_{\mathsf{M}}$ on $\mathsf{M}\equiv(\mathcal{D},|\omega|^2)$.

Formulas (\ref{derivatives}), (\ref{variation of resolvent kernel - formula}) and the equality $\overline{R_\lambda(x,y)}=R_{\overline{\lambda}}(x,y)$ imply
\begin{align*}
\frac{\partial R_\lambda(x,y)}{\partial t_k}=&\Re\Big(\frac{\partial R_\lambda(x,y)}{\partial C_{+k}}-\frac{\partial R_\lambda(x,y)}{\partial C_{-k}}\Big)=\\
=4\int\limits_{C_{+k}^\ddag-C_{-k}^\ddag}&\mathscr{W}_\lambda(x|\cdot|y)=4\Im\int\limits_{C_{+k}^\ddag}\big[\mathscr{W}_\lambda+\tau^*\mathscr{W}_\lambda\big](x|\cdot|y) \qquad (\lambda\in\mathbb{R}\backslash{\rm Sp}(\Delta_X)),
\end{align*}
where $C_{\pm k}=-\tau\circ C_{\pm k}$ is the clockwise contour enclosing $p_{\pm k}=\tau(p_{\pm k})$. In view of $\tau^*\partial f=\overline{\partial}(f\circ\tau)$ and (\ref{reslovent kernel symmetry}), we have $\tau^*\partial_z R_{\lambda}(z,y)=\partial_{\overline{z^\dag}}R_{\lambda}(z^\dag,\tau(y))$, where $z^\dag=\tau(z)$. Due to this fact, anti-symmetry (\ref{anti-symmetry of differential}) of $\omega$, and the equality $\overline{R_{\overline{\lambda}}(x,y)}=R_\lambda(x,y)=R_\lambda(y,x)$, we obtain
\begin{equation}
\label{antisimmetry of variation difform}
\tau^*\mathscr{W}_\lambda(x|\cdot|y)=-\overline{\mathscr{W}_{\overline{\lambda}}(\tau(x)|\cdot|\tau(y))}.
\end{equation}
Thus, we arrive at
\begin{equation}
\label{tvariation of RK on double}
\frac{\partial R_\lambda(x,y)}{\partial t_k}=4\Im\int\limits_{C_{+k}^\ddag}\big[\mathscr{W}_\lambda(x|\cdot|y)+\mathscr{W}_{\lambda}(\tau(x)|\cdot|\tau(y))\big] \qquad (\lambda\in\mathbb{R}\backslash{\rm Sp}(\Delta_X)).
\end{equation}
Combining (\ref{Dirichlet resolvet kernel}) and (\ref{tvariation of RK on double}) yields
\begin{align*}
\frac{\partial R^D_\lambda(x,y)}{\partial t_k}=4\Im\int\limits_{C_{+k}^\ddag}\sum_{\alpha,\beta=0}^{1}(-1)^{\alpha+\beta}\mathscr{W}_\lambda(\tau^{\alpha}x|\cdot|\tau^{\beta}y)
\end{align*}
From (\ref{Dirichlet resolvet kernel}), it easily follows that
\begin{align*}
\sum_{\alpha,\beta=0}^{1}(-1)^{\alpha+\beta}R_\lambda(\tau^\alpha(x),z)R_{\lambda}(z,\tau^{\beta}(y))=R^D_\lambda(x,z)R^D_{\lambda}(z,y).
\end{align*}
Combining the last two formulas with (\ref{variation of resolvent kernel - difform}), one arrives at
\begin{equation}
\label{tvariation of RK Dirihlet}
\frac{\partial R^D_\lambda(x,y)}{\partial t_k}=4\Im\int\limits_{C_{+k}^\ddag}\mathscr{W}^D_\lambda(x|\cdot|y) \qquad (\lambda\in\mathbb{R}\backslash{\rm Sp}(\Delta_{\mathsf{M}})),
\end{equation}
where
\begin{equation}
\label{variation of RK Dirichlet - difform}
\mathscr{W}^D_\lambda(x|z|y)=\frac{\partial_z R^D_\lambda(x,z)\partial_z R^D_{\lambda}(z,y)-(\lambda/4) R^D_\lambda(x,z)R^D_\lambda(z,y)|\omega(z)|^2}{\omega(z)}.
\end{equation}

Similarly, formulas (\ref{derivatives}), (\ref{variation of resolvent kernel - formula}), and (\ref{antisimmetry of variation difform}) lead to
\begin{align*}
\frac{\partial R_\lambda(x,y)}{\partial \delta_s}&=2\Im\Big[\frac{\partial R_\lambda(x,y)}{\partial B_{i(s)}}+\sum_{\pm k}\frac{\partial R_\lambda(x,y)}{\partial C_{\pm k}}\mathbbm{1}_\cap(\Pi_s,\mathfrak{P}_k)\Big]=\\
=&-4\Re\Bigg[\,\oint\limits_{a_{i(s)}}\mathscr{W}_\lambda(x|\cdot|y)-\oint\limits_{\sum_k \mathbbm{1}_\cap(\Pi_s,\Gamma_i) C_{+k}^\ddag}[\mathscr{W}_\lambda-\tau^*\mathscr{W}_\lambda](x|\cdot|y)\Bigg]=\\
=&4\Re\oint\limits_{\delta_s^\ddag}[\mathscr{W}_\lambda-\tau^*\mathscr{W}_\lambda](x|\cdot|y)=\\
=&4\Re\oint\limits_{\delta_s^\ddag}[\mathscr{W}_\lambda(x|\cdot|y)+\mathscr{W}_\lambda(\tau(x)|\cdot|\tau(y))] \qquad (\lambda\in\mathbb{R}\backslash{\rm Sp}(\Delta_X))
\end{align*}
and
\begin{equation}
\label{deltavariation of RK Dirihlet}
\frac{\partial R^D_\lambda(x,y)}{\partial\delta_s}=4\Re\int\limits_{\delta_s^\ddag}\mathscr{W}^D_\lambda(x|\cdot|y) \qquad (\lambda\in\mathbb{R}\backslash{\rm Sp}(\Delta_{\mathsf{M}})),
\end{equation}
where $i(s)$ is the unique number obeying $\Gamma_{i(s)}\equiv b_{i(s)}$ intersects the rectangle $\Pi_s$, the contour $-B_{i(s)}^\ddag=a_{i(s)}$ is represented as $a_{i(s)}=h_s-\tau\circ h_s$, where $h_s$ is the horizontal side $[0,1]\times\{0\}$ of the rectangle $\Pi_s$, and the notation
\begin{equation}
\label{dual contours}
t_k^\ddag:=C_{+k}^\ddag, \quad \delta_s^\ddag:=\sum_k \mathbbm{1}_\cap(\Pi_s,\Gamma_i) C_{+k}^\ddag-h_s
\end{equation}
is used.

\subsection{Variations of eigenvalues}
Let $0<\lambda_1^D(\sigma)\le\lambda_2^D(\sigma)\le\dots$ be the eigenvalues of the Laplacian $\Delta_{\mathsf{M}}=\Delta_{\mathsf{M}(\sigma)}$ on $\mathsf{M}=\mathsf{M}(\sigma)$ repeated according to their multiplicities and let $\{u_k^{D,(\sigma)}\}_{k=1}^\infty$ be an orthonormal basis of the corresponding eigenfunctions in $L_2(\mathsf{M})$. 
(Note that $(u,\lambda)$ is an eigenpair of $\Delta_{\mathsf{M}}$ if and only if $u$ can be extended to the eigenfunction of $\Delta_X$ on $X\supset\mathsf{M}$ corresponding to the eigenvalue $\lambda$ by the rule $u\circ\tau=-u$.)

Let $\sigma'$ be close to $\sigma$. The formula $R_{\lambda}^{D}(x,y)=\sum_{k=1}^{\infty}u_k^{D}(x)\overline{u_k^{D}(y)}/(\lambda_k^{D}-\lambda)$ implies
\begin{align}
\label{residue of resolvent}
-\frac{1}{2\pi i}\oint\limits_{\partial U}\dot{R}_{\lambda}^{D,(\sigma')}(x,y)f(\lambda)d\lambda\Bigg|_{y=x}=\frac{\partial}{\partial\sigma'}\Big[\sum_{\lambda_k^D(\sigma')\in U}f(\lambda_{k}^D(\sigma'))|u_k^{D,(\sigma')}(x)|^2\Big],
\end{align}
where $U$ is a domain whose boundary does not intersect ${\rm Sp}(\Delta_{\mathsf{M}(\sigma)})$ and $f$ is holomorphic on $\overline{U}$. Let $U$ contain exactly one eigenvalue $\lambda_j^{D}(\sigma)=\dots=\lambda_{j+m-1}^{D}(\sigma)$ of multiplicity $m$ and $f(\lambda)=\lambda-\lambda_j^D(\sigma)$. Under the assumption that $\lambda_k^D(\sigma)$ and $u_k^{D,(\sigma)}$ are differentiable in $\sigma$ separately, one can rewrite the right-hand side of (\ref{residue of resolvent}) as follows
$$\sum_{\lambda_k^D(\sigma)\in U}\dot{\lambda}_{k}^D(\sigma)|u_k^{D,(\sigma)}(x)|^2+o(1) \ (\sigma'\to\sigma).$$
Put $\sigma'=\sigma$ in (\ref{residue of resolvent}) and integrate over $\mathsf{M}$ (here we use the assumption that the asymptotics of $R_{\lambda}^{D,(\sigma')}(x,y)$ as $\sigma'\to\sigma$, provided by (\ref{tvariation of RK Dirihlet}), (\ref{deltavariation of RK Dirihlet}), is uniform for $x,y$ close to the contours near which the geometric surgery is performed). As a result one obtains the following expression
\begin{align}
\label{variation of eigenvalues general}
\dot{\Sigma}^D_j(\sigma):=-\int\limits_{\mathsf{M}}\underset{\lambda=\lambda_j^{D}(\sigma)}{\rm Res}\Big[\dot{R}_{\lambda}^{D,(\sigma)}(x,y)\big(\lambda-\lambda_j^D(\sigma)\big)\Big]\Bigg|_{y=x}dS(x)
\end{align}
for the sum $\Sigma^D_j(\sigma')$ of the eigenvalues of $\Delta_{\mathsf{M}(\sigma')}$ that are close to $\lambda_j^D(\sigma)$ as $\sigma'$ is close to $\sigma$. The justification of (\ref{variation of eigenvalues general}) without the above assumptions is made in \S 6.3, \cite{KoKori}. 
The residue in (\ref{variation of eigenvalues general}) can be found from the asymptotics of $\dot{R}_{\lambda}^{D,(\sigma)}$ for {\it real} $\lambda$ close to $\lambda_j^{D}(\sigma)$ provided by formulas (\ref{tvariation of RK Dirihlet}), (\ref{deltavariation of RK Dirihlet}), (\ref{variation of RK Dirichlet - difform}). Namely, the formulas
\begin{align*}
\underset{\lambda=\lambda_j^{D}}{\rm Res}\Big[\mathscr{W}^D_\lambda(x|\cdot|x)(\lambda-\lambda_j^{D})\Big]=\sum_{\lambda_k^{D}(\sigma)=\lambda_j^{D}}|u_k^{D}(x)|^2\mathscr{W}_{(k)}^D
\end{align*}
and
\begin{equation}
\label{difforms for eigenvalues}
\mathscr{W}_{(k)}^D:=\omega^{-1}(\partial u_k^{D})^2-(\lambda_k^{D}/4)\,\overline{\omega}(u_k^{D})^2
\end{equation}
(and also their analogues with omitted $D$) yield
\begin{align}
\label{variation of eigenvalues Dirichlet}
\frac{\partial \Sigma^D_j}{\partial t_k}=-4\Im\int\limits_{t_k^\ddag}\sum_{\lambda_l^{D}=\lambda_j^{D}}\mathscr{W}_{(l)}^D, \qquad \frac{\partial \Sigma^D_j}{\partial\delta_s}=-4\Re\int\limits_{\delta_s^\ddag}\sum_{\lambda_l^{D}=\lambda_j^{D}}\mathscr{W}_{(l)}^D,
\end{align}
where the contours $\delta_s^\ddag,t_k^\ddag$ are defined in (\ref{dual contours}).

\subsection{Variation of zeta functions}
Note that
$$\sum_{l=1}^{\infty}\frac{(\partial^q u^D_l(y))^2}{\lambda^D_l-\lambda}=\partial^q_x\partial^q_y R^D_{\lambda}(x,y)\big|_{x=y}, 
$$
where $q=0,1$. 
In view of the above equalities and formulas (\ref{variation of eigenvalues Dirichlet}),  
the zeta function $\zeta_{\Delta_{\mathsf{M}}-\lambda}(\varsigma):=\sum_{l=1}^{\infty}(\lambda_l^D-\lambda)^{-\varsigma}$ 
admits term-wise differentiation in the parameters $\delta_s$, $t_k$ for $\Re\varsigma$ sufficiently large, and the equalities 
\begin{align}
\label{zeta dervatives Dirichlet}
\begin{split}
\frac{\partial \zeta_{\Delta_{\mathsf{M}}-\lambda}(N)}{\partial t_k}&=\sum_{l=1}^{\infty}\frac{4N\,\Im\int_{t_k^\ddag}\mathscr{W}_{(l)}^D}{(\lambda_l^D-\lambda)^{N+1}}=-2i\int\limits_{t_k^\ddag}\frac{\partial_\lambda^{N}\big(\Omega^D_{\lambda}(x,y)-\overline{\Omega^D_{\overline{\lambda}}(x,y)}\big)}{(N-1)!}\Bigg|_{x=y},\\
\frac{\partial \zeta_{\Delta_{\mathsf{M}}-\lambda}(N)}{\partial \delta_s}&=2\int\limits_{\delta_s^\ddag}\frac{\partial_\lambda^{N}\big(\Omega^D_{\lambda}(x,y)+\overline{\Omega^D_{\overline{\lambda}}(x,y)}\big)}{(N-1)!}\Bigg|_{x=y},
\end{split}
\end{align}
hold for all sufficiently large $N\in\mathbb{N}$, where
\begin{equation}
\label{zeta variation difform}
\begin{split}
\Omega^D_{\lambda}(x,y):&=\omega^{-1}(y)\,\partial_{x}\partial_y R^D_\lambda(x,y)-(\lambda/4)\,\overline{\omega(x)} R^D_\lambda(x,y).
\end{split}
\end{equation}

\section{Variations of determinants of Laplacians}
\subsection{Regularization of diagonal values of $\Omega_\lambda$ and $\Omega^D_\lambda$}
Recall that the resolvent kernel $R_\lambda(x,y)$ admits expansion
\begin{align}
\label{resolvent kernel near diagonal}
\begin{split}
R_\lambda(\cdot,y)=\big(I-(\Delta_{X}&-\lambda)^{-1}(\Delta-\lambda)\big)\Big[\frac{\chi(r(\cdot,y))}{2\pi}K_0(r(\cdot,y)\cdot\sqrt{-\lambda})\Big]=\\
=&-\frac{\chi(r(\cdot,y))}{2\pi}{\rm log}(r(\cdot,y))\,\Big(1-\frac{\lambda r^2(\cdot,y)}{4}\Big)+\tilde{R}_\lambda(\cdot,y),
\end{split}
\end{align}
where $y$ is not a conical point on $(X,|\omega|^2)$, $r(x,y):={\rm dist}_{|\omega|^2}(x,y)$, $\chi$ is a cut-off function equal to 1 near the origin, ${\rm supp}\chi$ is sufficiently small, and the remainder obeys $\tilde{R}_\lambda\in C^3(X\times X)$. The last equality follows from the smoothness-increasing theorems for solutions to elliptic equations and the expansion
$$K_0(z)=-{\rm log}\Big(\frac{z}{2}\Big)\Big(1+\frac{z^2}{4}\Big)-\gamma-\frac{z^2}{4}\Big(\gamma-1\Big)+O(z^4{\rm log}(z)).$$

In view of (\ref{resolvent kernel near diagonal}) and (\ref{Dirichlet resolvet kernel}), the section 
$\Omega^D_{\lambda}(x,y)$ has singularity at the diagonal $x=y$. Therefore, we introduce the following regularizations with well-defined diagonal values,
\begin{equation}
\label{zeta variation difform regularized}
\begin{split}
\Omega^{D,(reg)}_{\lambda}(x,y):=\omega^{-1}(y)\partial_{x}\partial_y \Big(R^D_\lambda(x,y)+\frac{{\rm log}(r)}{2\pi}\,\Big(1-\frac{\lambda r^2}{4}\Big)\Big)-\\
-(\lambda/4)\,\overline{\omega(x)}\Big(R^D_\lambda(x,y)+\frac{{\rm log}(r)}{2\pi}\Big).
\end{split}
\end{equation}
where $r=r(x,y)$. Since the difference 
$\Omega^{D}_{\lambda}-\Omega^{D,(reg)}_{\lambda}$ is linear in $\lambda$, one can replace $\Omega^{D}_{\lambda}$ with $\Omega^{D,(reg)}_{\lambda}$ and interchange the integration and the differentiation in $\lambda$ in formulas (\ref{zeta dervatives Dirichlet}). Note that $\Omega^{D,(reg)}_{\lambda}(x,x)$ is smooth up to $\partial\mathsf{M}$ due to (\ref{Dirichlet resolvet kernel}) and the equality
\begin{equation}
\label{Mcdonald eq}
(4\partial_z\partial_{z'}-\lambda)K_0(\sqrt{-\lambda}|z-\overline{z'}|)=0.
\end{equation}
Introduce the meromorphic functions
\begin{align}
\label{F def}
\begin{split}
F(\lambda\,|\,t_k):&=-2i\int\limits_{t_k^\ddag}\big(\Omega^{D,(reg)}_{\lambda}(x,x)-\overline{\Omega^{D,(reg)}_{\overline{\lambda}}(x,x)}\big),\\
F(\lambda\,|\,\delta_s):&=2\int\limits_{\delta_s^\ddag}\big(\Omega^{D,(reg)}_{\lambda}(x,x)+\overline{\Omega^{D,(reg)}_{\overline{\lambda}}(x,x)}\big),
\end{split}
\end{align}
then formulas (\ref{zeta dervatives Dirichlet}) can be rewritten as
\begin{equation}
\label{zeta derivatives general}
(N-1)!\partial_{\diamond}\zeta_{\Delta_{\Delta_{\mathsf{M}}}-\lambda}(N)=\partial_{\lambda}^N F(\lambda\,|\,\diamond) \qquad (\diamond=t_k,\delta_s).
\end{equation}

\subsection{Values of $\Omega^{D,(reg)}_{\lambda}(x,x)$ at $\lambda=0$} Recall that the Green function $G^D(x,y)=R^D_0(x,y)$ of $\Delta^D$ is conformally invariant: if $\beta:\mathsf{M}\mapsto M$ is a smooth (up to the boundary) conformal map, then the Green function $\tilde{G}$ of the Dirichlet Laplacian on $M$ is related to $G^D$ via $G^D(x,y)=\tilde{G}(\beta(x),\beta(y))$. Thus, in the calculations involving $G^D(x,y)$, one can identify $\mathsf{M}=\mathfrak{A}([\mathcal{D}])$ with any domain $\mathcal{D}\in[\mathcal{D}]$ via the biholomorphism provided by Lemma \ref{domais vs reduced LC}. 

Introduce the {\it Schiffer kernel} $S$ of $\mathsf{M}\equiv\mathcal{D}$ by 
\begin{equation}
\label{Schiffer kernel}
S(x,y):=-4\pi\frac{\partial^2G^D(x,y)}{\partial x\,\partial y}
\end{equation}
(see \cite{BerSch}). Alternatively, $S$ is defined as the bimeromorphic section of $K_x\otimes K_y$ over $\mathsf{M}$ with the singularity at the diagonal
\begin{equation}
\label{Schiffer connection}
S(x,x')=\frac{1}{(x-x')^2}+\frac{\mathscr{S}(x)}{6}+o(1), \qquad |x'-x|\to 0
\end{equation}
(here $x,x'$ are different values of the same holomorphic coordinate). The {\it Schiffer projective connection} $\mathscr{S}$ transforms as
$$\mathscr{S}(y)=\mathscr{S}(x)\Big(\frac{\partial x}{\partial y}\Big)^2+\{x,y\}$$
under the holomorphic change of the coordinates (here $\{\cdot,\cdot\}$ is the Schwarzian derivative). Introduce also the projective connection 
\begin{equation}
\label{distinguished}
\mathcal{S}_\omega:=\{z,x\}, \qquad z(p):=\int\limits_\cdot^p\omega
\end{equation}
(note that the coordinate $z$ provides the local isometry between $\mathsf{M}\backslash\{p_1,\dots,p_{n-1}\}$ and $\mathbb{C}$). Then the difference $\mathscr{S}-\mathcal{S}_\omega$ is a quadratic differential.

Since $\omega(z)=dz$ and $r(x,y)=|z(x)-z(y)|$ for $x,y$ close to each other and separated from conical points, the substitution $\lambda=0$ into (\ref{zeta variation difform regularized}) yields $\Omega^{D,(reg)}_{0}(z,z)=-\mathscr{S}(z)/24\pi$, whence
\begin{equation}
\label{Omega 0 Dirichlet}
\Omega^{D,(reg)}_{0}(x,x)=-\frac{[\mathscr{S}-\mathcal{S}_\omega](x)}{24\pi\omega(x)}.
\end{equation}

In view of (\ref{Omega 0 Dirichlet}), formulas (\ref{F def}) imply
\begin{align}
\label{F 0}
\begin{split}
F(0\,|\,t_k)&=-\Im\int\limits_{t_k^\ddag}\frac{\mathscr{S}-\mathcal{S}_\omega}{6\pi\omega}, \quad F(0\,|\,\delta_s)=-\Re\int\limits_{\delta_s^\ddag}\frac{\mathscr{S}-\mathcal{S}_\omega}{6\pi\omega}.
\end{split}
\end{align}
\subsection{Asymptotics of $\Omega^{D,(reg)}_{\lambda}(x,x)$ as $\lambda\to -\infty$}
Since $K_0(z)$ and all its derivatives decay exponentially as $|z|\to \infty$, $|{\rm arg}(z)|<\pi$, formulas (\ref{resolvent kernel near diagonal}), (\ref{Dirichlet resolvet kernel}), the estimate $\|(\Delta_{X}-\lambda)^{-1}\|_{B(L_2(\mathsf{M}))}=O(|\lambda|^{-1})$ and the standard smoothness increasing theorems for solutions to elliptic equations provide the asymptotics
\begin{align*}
R_\lambda(x,y)&=\frac{\chi(r)}{2\pi}K_0(r\sqrt{-\lambda})+\tilde{R}_\lambda(x,y);\\
R^D_\lambda(x,y)&=\frac{\chi(r)}{2\pi}K_0(r\sqrt{-\lambda})-\frac{\chi(r^\dag)}{2\pi}K_0(r^\dag\sqrt{-\lambda})+\tilde{R}^D_\lambda(x,y),
\end{align*}
where $r={\rm dist}_{|\omega|^2}(x,y)$, $r^\dag:={\rm dist}_{|\omega|^2}(x,\tau(y))$ while the remainders $\tilde{R}_\lambda,\tilde{R}^D_\lambda$ and all their derivatives in $x,y$ decay exponentially as $\Re\lambda\to -\infty$ and uniformly in $x,y$ separated from conical points. Therefore, formulas (\ref{zeta variation difform regularized}) and (\ref{Mcdonald eq}) imply the asymptotics
\begin{equation}
\label{Omega infty}
\begin{split}
\Omega^{D,(reg)}_{\lambda}(z,z)&=\frac{\lambda\,d\overline{z}}{8\pi}\Big(\frac{{\rm log}(-\lambda/4)}{2}+\gamma\Big)+\frac{\lambda\,(d\overline{z}-dz)}{8\pi}K_0(2\mathfrak{d}\sqrt{-\lambda})+O(e^{-\epsilon\Re\sqrt{-\lambda}})
\end{split}
\end{equation}
as $\Re\lambda\to-\infty$, where $\epsilon>0$, $\mathfrak{d}:={\rm dist}_{|\omega|^2}(z,\partial\mathsf{M})$, and the coordinate $z$ is given by (\ref{distinguished}). Note that the last term in the right-hand side of the last formula in (\ref{Omega infty}) is zero on any tangent vector of the contour $h_s$; thus, its integral over any contour (\ref{dual contours}) is $O(e^{-\epsilon\Re\sqrt{-\lambda}})$. As a corollary, functions (\ref{F def}) admit the following asymptotics
\begin{align}
\label{F infty}
\begin{split}
F(\lambda\,|\,t_k)=O(e^{-\epsilon\Re\sqrt{-\lambda}}), \qquad 
F(\lambda\,|\,\delta_s)+\frac{\lambda}{2\pi}\Big(\frac{{\rm log}(-\lambda/4)}{2}+\gamma\Big)=O(e^{-\epsilon\Re\sqrt{-\lambda}}).
\end{split}
\end{align}
as $\Re\lambda\to-\infty$; where the left-hand sides can be differentiated in $\lambda$. 

\subsection*{Variations of ${\rm det}\Delta_{\mathsf{M}}$ and ${\rm det}'\Delta_{X}$}
In view of the residue theorem, we have
$$(\varsigma-1)\dots(\varsigma+1-N)\tilde{\lambda}^{-\varsigma}=\frac{(N-1)!}{2\pi i}\int_\Gamma\frac{\lambda^{N-1-\varsigma}d\lambda}{(\tilde{\lambda}-\lambda)^{N}},$$
where $\Gamma$ is the contour enclosing the cut $(-\infty,0]$. Summing over $\tilde{\lambda}\in{\rm Sp}(\Delta_{\mathsf{M}})$, we arrive at
$$(\varsigma-1)\dots(\varsigma-N+1)\zeta_{\Delta_{\mathsf{M}}}(\varsigma)=\frac{(N-1)!}{2\pi i}\int\limits_\Gamma\zeta_{\Delta_{\mathsf{M}}-\lambda}(N)\lambda^{N-1-\varsigma}d\lambda$$
If $\Re\varsigma\gg 1$, then both sides are well-defined and admit a meromorphic continuation onto the whole $\mathbb{C}$. Differentiation of both sides with respect to the parameters $\diamond=t_k,\delta_s$ for large positive $N$ and $\Re\varsigma$ and taking into account (\ref{zeta derivatives general}) yield
\begin{align*}
(\varsigma-1)\dots(\varsigma-N+1)\partial_{\diamond}\zeta_{\Delta_{\mathsf{M}}}(\varsigma)=\frac{1}{2\pi i}\int\limits_\Gamma\lambda^{N-1-\varsigma}\partial_{\lambda}^N F(\lambda\,|\,\diamond)d\lambda.
\end{align*}
Here the right-hand side admits a meromorphic continuation onto $\mathcal{C}\backslash\{1\}$. Since the integration in the parameter $\diamond$ can be interchanged with meromorphic continuation, formula (\ref{zeta s variation}) is valid for all $s\in\mathbb{C}$. Integrating by parts and taking into account (\ref{F infty}) one finally obtains
\begin{equation}
\label{zeta s variation}
\partial_{\diamond}\zeta_{\Delta_{\Delta_{\mathsf{M}}}}(\varsigma)=\frac{1}{2\pi i(\varsigma-1)}\int\limits_\Gamma\lambda^{1-\varsigma}\partial_{\lambda}^2 F(\lambda\,|\,\diamond)d\lambda.
\end{equation}
Now, we make use of the following lemma (see Lemma 3.1, \cite{KoKoPolyPoly}).
\begin{lemma}
\label{magiclemma}
Let $F$ be a function holomorphic in some neighborhood of $(-\infty,0]$ containing the curve $\Gamma$. Suppose that the asymptotics
\begin{equation}
\label{F asymp}
F(\lambda)=\sum_{k=1}^K (F_k+\tilde{F}_k\lambda{\rm log}(-\lambda))\lambda^{r_k}+\Phi(\lambda)
\end{equation}
is valid as $\Re\lambda\to -\infty$, where $r_k\in\mathbb{R}$, $F_k,\tilde{F_k}\in\mathbb{C}$, and $|\lambda^k\partial^k_\lambda\Phi(\lambda)|=O(\lambda^{\kappa})$ for some $\kappa<0$ and all $k=0,1,\dots$. Denote by $F(\infty)$ and $\tilde{F}(\infty)$ the constant term and the coefficient at ${\rm log}(-\lambda)$ in {\rm(\ref{F asymp})}. Let $\widehat{F}$ be the analytic continuation of the integral
$$\widehat{F}(\varsigma):=\int\limits_{\Gamma}\partial_\lambda^2F(\lambda)\,\frac{\lambda^{1-\varsigma}d\lambda}{2\pi i}.$$
Then $\widehat{F}$ is holomorphic at $\varsigma=0$ and
$$\widehat{F}(0)=\tilde{F}(\infty), \qquad \partial_\varsigma\widehat{F}(0)=F(\infty)-\tilde{F}(\infty)-F(0).$$
In particular, the function $\varsigma\mapsto\eta(\varsigma):=\widehat{F}(\varsigma)/(\varsigma-1)$ obeys
$$\eta(0)=-\tilde{F}(\infty), \qquad -\partial_\varsigma\eta(0)=F(\infty)-F(0).$$
\end{lemma}
Substituting $F=F(\cdot\,|\,\diamond)$ into Lemma \ref{magiclemma} and taking into account (\ref{zeta s variation}), one arrives at 
\begin{equation}
\nonumber
\partial_{\diamond}{\rm log}\,{\rm det}(\Delta_{\mathsf{M}})=-\partial_\sigma\partial_{\diamond}\zeta_{\Delta_{\mathsf{M}}}(0)=F(\infty\,|\,\diamond)-F(0\,|\,\diamond).
\end{equation}
In view of (\ref{F infty}), we have $F(\infty\,|\,\diamond)=0$, 
while $F(0\,|\,\diamond)$ is given by (\ref{F 0}). Hence, we finally obtain
\begin{equation}
\label{det variation final}
\partial_{t_k}{\rm log}\,{\rm det}(\Delta_{\mathsf{M}})=\Im\int\limits_{t_k^\ddag}\frac{\mathscr{S}-\mathcal{S}_\omega}{6\pi\omega}, \quad \partial_{\delta_s}{\rm log}\,{\rm det}(\Delta_{\mathsf{M}})=\Re\int\limits_{\delta_s^\ddag}\frac{\mathscr{S}-\mathcal{S}_\omega}{6\pi\omega}.
\end{equation}

\begin{rem} Note that ${\rm det\,}(\Delta_{\mathsf{M}})$ is related to the determinant, ${\rm det\,}\Delta_{\mathcal D}$, of the operator of the Dirichlet boundary problem in $\mathcal {D}$ through the following analogue of the Alvarez-Polyakov formula for flat surfaces with conical singularities 
(which is essentially a minor modification of the CHS (Carron-Hillairet-Spreafico) formula from \cite{KokPAMS} for flat conical surfaces without boundary). 

Let $z_k$ ($k=1, \dots, n-1$) be the coordinates of the zeroes, $P_k$, of the holomorphic one-form $\omega$ in $\mathcal{D}$ and let $\omega=(z-z_k)\omega_k(z)dz$ near $P_k$. Let also $\omega=\omega(z)dz$ in $\mathcal {D}$ outside the zeroes $P_k$.  

Then
\begin{equation}\label{CHS}
\log\frac{{\rm det}\,\Delta_{\mathsf{M}}} {{\rm det\,}\Delta_{\mathcal D}   }=-\frac{1}{12\pi}\int_{\partial \mathcal{D}}\log |\omega(z)|\partial_\nu\log |\omega(z)|dl(z)
\end{equation}
$$
-\frac{1}{6\pi}\int_{\partial \mathcal{D}}k(z)\log |\omega(z)|dl(z)+\frac{1}{12}\sum_{k=1}^{n-1}\log
|\omega_k(z_k)|+d(n)\,$$ 
where $k$ is the geodesic curvature and $d(n)$ is an absolute constant depending only on $n$. 
\end{rem}

\section{Variations of determinant of DN map}
Let $(X,g)$ be a smooth Riemannian surface divided into two parts $\mathsf{M}$, $\tilde{\mathsf{M}}$ by smooth curve $\Gamma$. Let $\Delta_X$ be the Laplacian on $X$ and let $\Delta_{\mathsf{M}}$ and $\Delta_{\tilde{\mathsf{M}}}$ be the Dirichlet Laplacians on $\mathsf{M}$, $\tilde{\mathsf{M}}$, respectively. Let $\Lambda:\,f\mapsto \partial_{\nu}u^f|_\Gamma$ and $\tilde{\Lambda}:\,f\mapsto \partial_{\tilde{\nu}}\tilde{u}^f|_\Gamma$ be the DN maps of $\mathsf{M}$ and $\tilde{\mathsf{M}}$, respectively, where $u^f$ and $\tilde{u}^f$ are harmonic extensions of $f\in H^1(\Gamma)$ into $\mathsf{M}$ and $\tilde{\mathsf{M}}$, respectively, and $\nu=-\tilde{\nu}$ is a normal vector on $\Gamma$ external to $\mathsf{M}$. The well-known BFK gluing formula (see Theorem $B^*$, \cite{BFK}, see also \cite{YLee}) reads
\begin{equation}
\label{BFK formula}
\frac{{\rm det}'(\Delta_X)}{{\rm det}(\Delta_{\mathsf{M}})\,{\rm det}(\Delta_{\tilde{\mathsf{M}}})}={\rm det}'(\Lambda+\tilde{\Lambda})\cdot \frac{{\rm Area}(X,g)}{|\Gamma|_g}.
\end{equation}
If $X$ is the double of the diagram $\mathsf{M}$, then ${\rm Area}(X,|\omega|^2)=|\partial\mathsf{M}|$ and (\ref{BFK formula}) takes the form
\begin{equation}
\label{BFK formula double}
{\rm det}'(\Lambda)=2{\rm det}'(2\Lambda)=\frac{2{\rm det}'(\Delta_X)}{({\rm det}(\Delta_{\mathsf{M}}))^2}
\end{equation}
(Here the following trick was used to calculate ${\rm det}'(2\Lambda)$. As is clear from the definition of the DN map $\Lambda_g$ of a Riemannian surface $(\mathsf{M},g)$, it transforms as $\Lambda_{\rho^2 g}=\rho^{-1}\Lambda_{g}$ under a conformal change  $g\mapsto\rho^2 g$ of the metric. At the same time, it is well-known that ${\rm det}'(\Lambda_g)/|\Gamma|_g$ is a conformal invariant, where $|\Gamma|_g$ is the length of $\Gamma$ in the metric $g$. Hence, $\zeta_{\Lambda_g}(0)=-1$ and ${\rm det}'(2\Lambda_g)={\rm det}'(\Lambda_{g/4})={\rm det}'(\Lambda_g)/2$.)

The variation of ${\rm det}'(\Delta_X)$ with respect to the parameters $\sigma=A_i,B_j$ is derived in Theorem 9, \cite{KoKorot},
\begin{align}
\label{Kokorot result}
\partial_\sigma{\rm log}\,\Big(\frac{{\rm det}'(\Delta_X)}{{\rm Area}(X,|\omega|^2)\,{\rm det}(\Im\mathbb{B})}\Big)=\frac{1}{12\pi i}\oint_{\sigma^\ddag}\frac{\mathcal{S}_B-\mathcal{S}_\omega}{\omega}.
\end{align}
Here $\mathcal{S}_B$ is the Bergman projective connection. If $X$ is the double of the diagram $\mathsf{M}$, then the canonical meromorphic bidifferential ${\bf  w}$ and the $\mathcal{S}_B$ obey the following symmetry conditions
$${\bf  w}(x^\dag,y^\dag)dxdy=\overline{{\bf  w}(x,y)}dx^\dag dy^\dag, \qquad \mathcal{S}_B(x^\dag)=\overline{\mathcal{S}_B(x)},$$
where $x,y$ are holomorphic coordinates of points $P,Q$ and $x^\dag=\overline{x},y^\dag=\overline{y}$ are holomorphic coordinates of $\tau(P)$, $\tau(Q)$, respectively. Thus, taking into account formulas (\ref{derivatives}), (\ref{dual contours}), (\ref{anti-symmetry of differential}), (\ref{periods via capacities}) and the equality ${\rm Area}(X,|\omega|^2)=2{\rm Area}(\mathsf{M})=2|\Gamma_0|_{m}=|\partial\mathsf{M}|$ one can rewrite (\ref{Kokorot result}) as 
\begin{align}
\label{Kokorot result appl}
\begin{split}
\partial_{t_k}{\rm log}\,\Big(\frac{{\rm det}'(\Delta_X)}{|\partial\mathsf{M}|\,{\rm det}(\mathscr{C})}\Big)=\Im\oint_{t_k^\ddag}\frac{\mathcal{S}_B-\mathcal{S}_\omega}{3\pi\omega};\\
\partial_{\delta_s}{\rm log}\,\Big(\frac{{\rm det}'(\Delta_X)}{|\partial\mathsf{M}|\,{\rm det}(\mathscr{C})}\Big)=\Re\oint_{\delta_s^\ddag}\frac{\mathcal{S}_B-\mathcal{S}_\omega}{3\pi\omega}.
\end{split}
\end{align}
Now combining (\ref{det variation final}) with (\ref{BFK formula double}) and (\ref{Kokorot result appl}) yields
\begin{equation}
\label{DET DN formula preliminary}
\begin{split}
\partial_{t_k}{\rm log}\Big(\frac{{\rm det}'(\Lambda)}{|\partial\mathsf{M}|\,{\rm det}(\mathscr{C})}\Big)=\Im\oint_{t_k^\ddag}\frac{\mathcal{S}_B-\mathscr{S}}{3\pi\omega},\\
\partial_{\delta_s}{\rm log}\Big(\frac{{\rm det}'(\Lambda)}{|\partial\mathsf{M}|\,{\rm det}(\mathscr{C})}\Big)=\Re\oint_{\delta_s^\ddag}\frac{\mathcal{S}_B-\mathscr{S}}{3\pi\omega}.
\end{split}
\end{equation}

For the Green function $G$ of $\Delta_X$, the following formulas are valid (see \cite{Fay})
\begin{align}
\label{Bergman proj connection}
-4\pi\frac{\partial^2 G(x,y)}{\partial x \partial y}&=\frac{1}{(x-y)^2}+\frac{S_B(y)}{6}-\pi\sum_{\alpha,\beta=1}^n(\Im\mathbb{B})^{-1}_{\alpha,\beta}\omega_\alpha(y)\omega_\beta(y)+o(1),\\
\label{Bergman kernel}
-4\pi\frac{\partial^2 G(x,y)}{\partial x \partial \overline{y}}&=:B(x,y)=\pi\sum_{\alpha,\beta=1}^n(\Im\mathbb{B})^{-1}_{\alpha,\beta}\omega_\alpha(y)\overline{\omega_\beta(y)}
\end{align}
(here $B$ is the Bergman kernel of $X$). In view of (\ref{Dirichlet resolvet kernel}), we have $G^D(x,y)=G(x,y)-G(x,\tau(y))$. Choose the local holomorphic coordinate $y^\dag$ on $X$ in such a way that $y^\dag(\tau(Q))=\overline{y(Q)}$. Recall that holomorphic differentials $\omega_\alpha$ are antisymmetric with respect to the involution on $X$, $\tau^*\omega_\alpha=-\overline{\omega_\alpha}$, i.e. 
$$\overline{\omega(y^\dag)}=-\omega(y)$$ (indeed, since $a$-cycles obey $a_k\circ\tau=-a_k$, the differentials $\overline{\tau^*\omega_\alpha}$ and $-\omega_\alpha$ are both holomorphic and have the same $a$-periods). Formulas (\ref{Bergman proj connection}) and (\ref{Bergman kernel}) imply
\begin{align*}
-\frac{1}{(x-y)^2}-4\pi\frac{\partial^2 G^D(x,y)}{\partial x \partial y}=-\frac{1}{(x-y)^2}-4\pi\frac{\partial^2 G(x,y)}{\partial x \partial y}+4\pi\frac{\partial^2 G(x,y^\dag)}{\partial x \partial\overline{y^\dag}}=\\
=\frac{S_B(y)}{6}-\pi\sum_{\alpha,\beta=1}^n(\Im\mathbb{B})^{-1}_{\alpha,\beta}\omega_\alpha(y)\Big(\omega_\beta(y)+\overline{\omega_\beta(y^\dag)}\Big)+o(1)=\frac{S_B(y)}{6}+o(1).
\end{align*}
Comparing this equality with (\ref{Schiffer kernel}), (\ref{Schiffer connection}), one obtains
\begin{equation}
\label{Shifer equal Bergman}
\mathscr{S}=S_B \text{ on } \mathsf{M}.
\end{equation}
Hence, the right-hand side of (\ref{DET DN formula preliminary}) vanishes and integration of (\ref{DET DN formula preliminary}) yields
\begin{equation}
\label{DET DN FINAL}
\frac{{\rm det}'(\Lambda)}{|\partial\mathcal{D}|}=c(n){\rm det}(\mathscr{C}),
\end{equation}
where $\Lambda=\Lambda_g$ is the DN map of arbitrary domain $\mathcal{D}\subset[\mathcal{D}]$ endowed with arbitrary conformal metric $g$ and $|\partial\mathcal{D}|=|\partial\mathcal{D}|_g$ is the length of the boundary in this metric; note that the value ${\rm det}'(\Lambda_g)/|\partial\mathcal{D}|_g$ is a conformal invariant. The constant $c(n)$ depends only on the number $n+1$ of connected components of $\partial\mathcal{D}$. (To ensure that $c(n)$ is independent of the top-dimensional cell containing $[\mathcal{D}]$, it is sufficient to note that all the steps of Subsections 3,4 can be repeated for the deformations leading to change of the NC-partition of $\mathcal{D}$ and shown on Fig. \ref{treordering}; due to this fact, the determinants ${\rm det}\,\Delta_\mathsf{M}$, ${\rm det}'\,\Delta_{X}$ and ${\rm det}'\,\Lambda$ are continuous functions on the whole stratum $\acute{\mathfrak{D}}_n$.)

For $n=1$, formula (\ref{DET DN FINAL}) reduces to the Guillarmou-Guillop\'e's formula 
\begin{equation}
\label{GG fomula ann}
\frac{{\rm det}'(\Lambda)}{|\partial\mathcal{D}|}=\frac{l(\gamma)}{\pi} \qquad (n=1)
\end{equation}
(see Theorem 1.2, \cite{Guillarmou}). Here $l(\gamma)$ is the length of the unique closed geodesic on the infinite cylinder $(\Pi:=\mathbb{R}\times(\mathbb{R}_r/\mathbb{Z}_t),h)$ equipped with the complete hyperbolic metric $h=dr^2+l(\gamma)^2{\rm cosh}^2(r)dt^2$, which is conformally equivalent to $\mathcal{D}$. Passing to the (flat) conformal metric $g=\pi^{-2}[dx^2+l(\gamma)^2dt^2]=\pi^{-2}{\rm cosh}^{-2}(r)h$ and taking into account that $\int_{-\infty}^{+\infty}\frac{dr}{{\rm cosh}(r)}=\pi$, one concludes that $\mathcal{D}$ is conformally equivalent to a cylinder $[0,1]\times(\mathbb{R}/\mathbb{Z}l(\gamma)\pi^{-1})=\mathsf{M}$ (endowed with the Euclidean metric). Thus, $\frac{l(\gamma)}{\pi}=|\partial\mathcal{D}_{ext}|_{m}=\int_{\partial\mathcal{D}_{ext}}2\partial u_0=\mathcal{C}$ due to (\ref{capacitances}) and the equality $m=|2\partial u_0|^2$. Now comparing (\ref{DET DN FINAL}) with (\ref{GG fomula ann}) yields $c(1)=1$.

To find the constant $c(n)$ in the general case, we consider the asymptotics of both sides of (\ref{DET DN FINAL}) as the diameters of all internal holes in $\mathcal{D}$ tend to zero. Namely, let $\mathcal{D}_\varepsilon=\{z\in\mathbb{C} \ | \ |z|\le 1, |z-z_j|\ge\varepsilon \ (j=1,\dots,n)\}$, where $|z_j|<1$ be a disk with $n$ small holes and let $\Lambda_\varepsilon$ be its DN map. The following asymptotic formula for ${\rm det}'(\Lambda_\varepsilon)$ is derived in Theorem 3.1, \cite{Wentworth}:
\begin{equation}
\label{Wentworth asymp}
\frac{{\rm det}'(\Lambda_\varepsilon)}{|\partial\mathcal{D}_\varepsilon|}=\Big(\frac{2\pi}{|{\rm log}(\varepsilon)|}\Big)^n(1+o(1)).
\end{equation}

Let $\mathcal{C}(\varepsilon)$ be the capacitance matrix of $\mathcal{D}_\varepsilon$. By definition, we have $\mathcal{C}_{ij}(\varepsilon)=\int_{\Gamma_i}\partial_\nu u_j dl$, where $\Gamma_i=\{z\in\mathbb{C} \ | \ |z-z_i|=\varepsilon\}$ ($i=1,\dots,n$), and $u_j$ is a harmonic function equal to $1$ on $\Gamma_j$. Let us represent $u_j$ as
$$u_j=\sum_{k=1}^n A_{jk}G(\cdot,z_k)+\tilde{u}_j,$$
where $G$ is the Green function of the Laplacian on the unit disk $\overline{\mathcal{D}_0}=\mathbb{D}$. Since each $G(\cdot,x_j)$ is harmonic in $\mathcal{D}_\varepsilon$ and obeys the asymptotics
$$G(\cdot,z_k)=-\frac{\delta_{ki}}{2\pi}{\rm log}|z-z_k|+\mathcal{M}_{ki}+O(|z-z_k|)$$
as $z\to z_i$, the remainder $\tilde{u}_j$ is harmonic in $\mathcal{D}_\varepsilon$ and obeys 
$$\tilde{u}_j|_{\Gamma_i}=\delta_{ji}-\big[A\Big(\frac{|{\rm log}\varepsilon|}{2\pi}I+\mathcal{M}\Big)\big]_{ji}+O(\varepsilon).$$ 
Choosing 
$$A=\Big(\frac{|{\rm log}\varepsilon|}{2\pi}I+\mathcal{M}\Big)^{-1},$$
one ensures that the remainder obeys $\tilde{u}_j(z)=O(\varepsilon)$ uniformly on $\mathcal{D}_\varepsilon$ due to the maximum principle. Let $\chi_i$ be the smooth cut-off function equal to one near $x_i$ and the support of $\chi_i$ is sufficiently small. Then the Green formula and the harmonicity of $\tilde{u}_j$ imply
\begin{align*}
\int_{\Gamma_i}\partial_\nu \tilde{u}_j\,dl=\int_{\Gamma_i}\partial_\nu \tilde{u}_j\,\chi_i\,dl-\int_{\Gamma_i}\tilde{u}_j\,\partial_\nu \chi_i\,dl=\int_{\mathcal{D}_\varepsilon}\tilde{u}_j\,\Delta_e\chi_i\,dS=O(\varepsilon)
\end{align*}
(here $\Delta_e=-4\partial_z\partial_{\overline{z}}$). Hence, we have
\begin{align*}
\mathcal{C}_{ij}(\varepsilon)=\sum_{k=1}^n A_{jk}\int_{\Gamma_i}\partial_\nu G(\cdot,x_k)+O(\varepsilon)=A_{ji}+O(\varepsilon)=\frac{2\pi}{|{\rm log}\varepsilon|}\delta_{ji}+O(|{\rm log}\varepsilon|^{-2})
\end{align*}
and
$${\rm det}(\mathcal{C}(\varepsilon))=\Big(\frac{2\pi}{|{\rm log}(\varepsilon)|}\Big)^n(1+o(1)).$$
Comparing the asymptotics of the two sides of (\ref{DET DN FINAL}) provided by the last formula and (\ref{Wentworth asymp}), one arrives at the equality
\begin{equation}
\label{ConsTtanT}
c(n)=1 \qquad (n=1,2,\dots).
\end{equation}

\begin{rem} Using (\ref{MAIN}), the BFK formula (\ref{BFK formula}), the CHS formula (\ref{CHS}) and, finally, the formula (1.10) from \cite{KoKorot} (the latter formula expresses  the determinant of the Laplacian on $X$ in the metric $|\omega|^2$ through the Bergman tau-function on the space of holomorphic differentials),
one gets the following expression for the determinant of the Dirichlet Laplacian in a multiply connected domain $\mathcal{D}$:

\begin{equation}\label{detDirD}
{\rm det}\,\Delta_{\mathcal D}=A(n)|\tau_B(X, \omega)|\prod_{j=1}^{n-1}|\omega_j(z_j)|^{-1/12}\exp\left\{I_1+I_2 \right\},
\end{equation}
where 
$$I_1=\frac{1}{12\pi}\int_{\partial \mathcal{D}}\log|\omega|\partial_\nu \log|\omega|\,dl\,,$$
$$I_2=\frac{1}{6\pi}\int_{\partial \mathcal {D}}k\log |\omega|\,dl\,,$$
$\tau(X, \omega)$ is the Bergman tau-function given by (3.24) from \cite{KoKorot}, and the constant $A(n)$ depends only on $n$.  

\end{rem}

\end{document}